\documentclass[prb,aps,amsmath,amsfonts]{revtex4}

\usepackage{graphicx}
\usepackage{hyperref}
\usepackage{multirow}
\usepackage{amssymb}

\def\sgn{\,\mbox{sgn}\,}

\newcommand{\ee}{\end{equation}}
\newcommand{\be}{\begin{equation}}
\newcommand{\la}{\label}
\newcommand{\bea}{\begin{eqnarray}}
\newcommand{\eea}{\end{eqnarray}}
\newcommand{\ba}{\begin{eqnarray}}
\newcommand{\ea}{\end{eqnarray}}

\newcommand{\ii}{{\rm i}}
\newcommand{\sign}{{\rm sgn}}
\newcommand{\de}{{\displaystyle\rm\mathstrut d}}

\def\XXint#1#2#3{{\setbox0=\hbox{$#1{#2#3}{\int}$}
     \vcenter{\hbox{$#2#3$}}\kern-.5\wd0}}

\begin{document}

\title{Nonlinear dynamics of spin and charge in spin-Calogero model}

\author{Manas Kulkarni$^{1,3}$, Fabio Franchini$^{2}$, Alexander G. Abanov$^{1}$}
\affiliation{$^{1}$Department of Physics and Astronomy, Stony Brook University, Stony Brook, NY 11794-3800}
\affiliation{$^{2}$The Abdus Salam ICTP; Strada Costiera 11, Trieste, 34100,
Italy}
\affiliation{$^{3}$Department of Condensed Matter Physics and Materials Science, Brookhaven National Laboratory, Upton, NY-11973}

\begin{abstract}
The fully nonlinear dynamics of spin and charge in spin-Calogero model is studied. The latter is an integrable one-dimensional model of quantum spin-1/2 particles interacting through inverse-square interaction and exchange. Classical hydrodynamic equations of motion are written for this model in the regime where gradient corrections to the exact hydrodynamic formulation of the theory may be neglected. In this approximation variables separate in terms of dressed Fermi momenta of the model. Hydrodynamic equations reduce to a set of decoupled Riemann-Hopf (or inviscid Burgers') equations for the dressed Fermi momenta. We study the dynamics of some non-equilibrium spin-charge configurations for times smaller than the time-scale of the gradient catastrophe. We find an interesting interplay between spin and charge degrees of freedom. In the limit of large coupling constant the hydrodynamics reduces to the spin hydrodynamics of the Haldane-Shastry model.
\end{abstract}
\maketitle

\tableofcontents

\section{Introduction}

One-dimensional models of many body systems have been a subject of intensive research since the seventies. Due to the low dimensionality, standard perturbative approaches developed in many body theory are often inapplicable. On the other hand some techniques specific to one spacial dimension are available and allow to treat systems of interacting particles non-perturbatively. The Fermi Liquid paradigm is replaced by the Luttinger Liquid theory \cite{1981-Haldane_JPC} in one dimension. One of its most striking predictions is that at low energies spin and charge degrees of freedom decouple. One can say that at low energies physical electrons exist as separate spin and charge excitations. At higher energies it is expected that spin and charge recombine into the original electrons. One can see the traces of spin-charge interaction taking into account corrections to the Luttinger liquid model arising from the finite curvature of band dispersion at Fermi energy \cite{1981-Haldane_JPC}. The coupling between spin and charge in one-dimensional systems was studied both perturbatively and using integrable models available in one dimension \cite{1994-BrazovskiiEtAl}.

In this paper we study the interaction between spin and charge in another integrable model -- the spin-Calogero model (sCM). This model is a spin generalization \cite{1992-Polychronakos-exchange,1992-HaHaldane,1993-HikamiWadati} of the well-known Calogero-Sutherland model
\cite{Sutherland-book}.

Calogero-Sutherland type models occupy a special place in 1D quantum physics. They are exactly solvable (integrable) but are very special even in the family of integrable models. In particular, they can be interpreted as systems of ``non-interacting'' particles with fractional exclusion statistics \cite{Sutherland-book,1991-Haldane-exclusion,1995-FukuiKawakami,1996-KatoKuramoto,1999-Polychronakos-LesHouches,1997-KatoYamamotoArikawa}.

The sCM model is given by the following Hamiltonian:
\begin{equation}
   H \equiv
   - \frac{\hbar^{2}}{2} \sum_{j=1}^{N}\frac{\partial^{2}}{\partial x_{j}^{2}}
   + \frac{\hbar^{2}}{2} \sum_{j\neq l} \frac{\lambda(\lambda\pm {\bf P}_{jl})} {\left(x_{j}-x_{l}\right)^2}
   \label{eq:general_H}
\end{equation}
where we took the mass of particles as a unity and ${\bf P}_{jl}$ is the operator that exchanges the positions of particles
$j$ and $l$ \cite{1992-Polychronakos-exchange}. The $\pm$ sign in the exchange term corresponds to ferromagnetic and anti-ferromagnetic ground state respectively if we are studying fermions. Similarly, it corresponds to anti-ferromagnetic and ferromagnetic ground state respectively if we are considering bosonic particles. The four cases can be summarized as:
\begin{eqnarray*}
   \mbox{Bosons} & \longrightarrow & \left\{ \begin{array}{ccl}
   + & \Rightarrow & \mbox{Anti-ferromagnetic} \; , \\
   - & \Rightarrow & \mbox{Ferromagnetic} \; ,\\ \end{array} \right. \\
   \mbox{Fermions} & \longrightarrow & \left\{ \begin{array}{ccl}
   + & \Rightarrow & \mbox{Ferromagnetic} \; ,\\
   - & \Rightarrow & \mbox{Anti-ferromagnetic} \; .\\ \end{array} \right.
\end{eqnarray*}
The coupling parameter $\lambda$ is positive and $N$ is the total number of particles.

As it has been already noted above the sCM is a very special model. In particular, in contrast to more generic integrable or non-integrable models the spin and charge in sCM are not truly separated even at low energies \cite{1992-HaHaldane}. Of course, one can still describe the low-energy excitation spectrum of sCM by two independent harmonic fluid Hamiltonians, one for the charge and the other for spin. However, it turns out that for the sCM the spin and charge velocities are the same \cite{1992-HaHaldane}, i.e. spin and charge do not actually separate.

Here we study the spin-Calogero model in the limit of an infinite number of particles using the hydrodynamic approach. Albeit, the collective field theory/quantum hydrodynamics of the spinless Calogero-Sutherland model has been studied in great detail \cite{1995-AwataEtAl,1983-AndricJevickiLevine,1988-AndricBardek,1995-Polychronakos,
2005-AbanovWiegmann}, a complete understanding of its spin generalization is still lacking although a considerable progress has been done recently in Refs. \cite{1996-AMY,jevicki-notes}.

We study the nonlinear collective dynamics of sCM in the semiclassical approximation, additionally neglecting gradient corrections to the equations of motion. This limit is justified as long as we consider configurations with small gradients of density and velocity fields. The gradientless approximation is commonly employed in studying nonlinear equations \cite{whithambook} and allows to study the evolution for a finite time while the first nonlinear contributions are dominant. For longer times, the solution will inevitably evolve toward configurations with large field gradients (such as shock waves) and the gradientless approximation becomes inapplicable. Nevertheless, in the initial stage of the evolution, corrections due to gradient terms in the equations of motion can be neglected (see further discussion in the Sec. \ref{ffRE}). We derive the gradientless hydrodynamics Hamiltonian from the Bethe Ansatz solution of the model.

The paper is organized as follows. In Sec.~\ref{sec:freeferm} we start with the simplest spinful integrable model -- a system of free fermions with spin. We briefly review the Bethe Ansatz solution for spin-Calogero model in Sec.~\ref{sec:sCM} and deduce the hydrodynamic Hamiltonian for the sCM from this solution in Sec.~\ref{sec:hydr} neglecting gradient corrections. The corresponding classical equations of motion are given in Sec.~\ref{sec:eqm}. It is shown that variables separate and the system of hydrodynamic equations is decoupled into four independent Riemann-Hopf equations for a given special linear combinations of density and velocity fields -- the dressed Fermi momenta. In Sec.~\ref{sec:Lattice} we illustrate that in the limit of strong coupling the hydrodynamics of sCM is reduced to the hydrodynamics of Haldane-Shastry lattice spin model giving the hydrodynamic formulation of the so-called {\it freezing trick} \cite{1993-Polychronakos}. We present some particular solutions of the hydrodynamic equations demonstrating nonlinear coupling between spin and charge degrees of freedom in the sCM in Sec.~\ref{sec:pictures} and conclude in Sec.~\ref{sec:concl}. To avoid interruptions in the main part of the paper some important technical details are moved to the appendices and are organized as follows.
In Appendix \ref{app:AABsCM} we use asymptotic Bethe ansatz to derive the hydrodynamics of sCM and to explain why variables separate in this system. In Appendix \ref{app:velocities} we describe the notion of true hydrodynamic velocities.  In Appendix \ref{app:allcases} we relate the hydrodynamics of sCM to two infinite families of mutually commuting conserved quantities and collect our results for the hydrodynamics in the different regimes of sCM. Finally, in Appendix \ref{app:hdHSM} we derive a hydrodynamic description of the Haldane-Shastry model from its Bethe Ansatz solution.

\section{Free fermions with spin}
 \la{sec:freeferm}

For one-dimensional free fermions without internal degrees of freedom the lowest state with a given total number of particles and total momentum corresponds to all single-particle plane wave states filled if the corresponding momentum $k$ satisfies $k_{L}<k<k_{R}$. Here $k_{L,R}$ are left and right Fermi momenta respectively which are defined by the given number of particles and momentum of the system:
\bea
   N/L & = &  \int_{k_L}^{k_R} \frac{\de k}{2\pi} =
    \frac{k_R - k_L}{2 \pi} = \rho  \; ,
   \label{rhoFF} \\
   P/L & = &  \int_{k_L}^{k_R} \frac{\de k}{2\pi} \; \hbar k
   = \hbar \frac{k^2_R - k^2_L}{4 \pi} = \rho v \; .
   \label{JFF}
\eea
Here we introduced the (overall) velocity of the system $v$ which is given from (\ref{rhoFF},\ref{JFF}) by
\be
   v/\hbar = \frac{k_R + k_L}{2} \; .
   \label{vFF}
\ee
Inverting (\ref{rhoFF},\ref{vFF}) we express the left and right Fermi points $k_{L,R}$ in terms of the density $\rho$ and velocity $v$ as
\be
  k_{R,L}  = v/\hbar \pm  \pi  \rho .
  \label{kRLFF}
\ee
The energy of this state is given by
\be
   E/L= \int_{k_L}^{k_R} \frac{\de k}{2\pi} \; \frac{\hbar^{2}k^2}{2}
   = \hbar^{2} \frac{k^3_R - k^3_L}{12 \pi}
   = \frac{\rho v^{2}}{2} + \frac{\hbar^{2}\pi^{2}}{6}\rho^{3}.
   \label{Hk3}
\ee
Up to this moment $\rho, v, k_{R,L}$ are just numbers characterizing the chosen state of free fermions (only two of them are independent). Assuming the locality of the theory we promote these numbers to quantum fields and write the hydrodynamic Hamiltonian of free spinless fermions as
\be
   {\cal H} = \int \de x\, \left[\frac{\rho(x) v^2 (x)}{2} + \frac{\hbar^2 \pi^2}{6} \rho^3 (x)\right]
   = \int \de x\,\hbar^{2}\frac{\left[ k_{R}(x) \right]^{3} - \left[ k_{L}(x) \right]^{3}}{12\pi}.
   \label{HFF}
\ee
Here we consider $\rho(x)$ and $v(x)$ as quantum field operators of density and velocity (and $k_{R,L}$ as given by (\ref{kRLFF})) having canonical commutation relations \cite{landau1941}
\be
   \left[ \rho(x), v(y) \right] = -\ii \hbar \delta' (x-y) \; .
 \la{rhov-comm}
\ee
Of course, gradient corrections to (\ref{HFF}) are generically present and the above ``derivation'' is just a heuristic argument (semiclassical in nature). It turns out that (\ref{HFF}) is, in fact, exact for free fermions.\footnote{It is exact if the nonlinear terms in (\ref{HFF}) are properly normal ordered.} It can be derived rigorously either using the method of collective field theory \cite{JevickiSakita,Sakita-book,Jevicki-1992} or conventional bosonization technique (but without linearization at Fermi points)\cite{1968-Schick,1981-Haldane_JPC,Schmeltzer-PRB-1993}.

The two terms of (\ref{HFF}) have a very clear physical interpretation. The first term is the kinetic energy of a fluid moving as a whole -- the only velocity term allowed by Galilean invariance.  The second one is the kinetic energy of the internal motion of particles. This term is finite due to the Pauli exclusion principle. Within the hydrodynamic approach we have to think of this term as of an internal energy of the fluid.

Commuting the Hamiltonian (\ref{HFF}) with the density and velocity operators one obtains the continuity and the Euler equations of quantum hydrodynamics. Alternatively, using $\left[ k_L (x) , k_L (y) \right] = - \left[ k_R (x) , k_R (y) \right] = 2\pi i \delta' (x-y)$ the equations of motion can also be written as a system of quantum Riemann-Hopf equations
\be
  \dot{k}_{R,L} + \hbar \; k_{R,L} \; \partial_x k_{R,L} = 0 \; .
  \label{FFRiemann}
\ee

For free fermions with spin, we simply add the Hamiltonians (\ref{HFF}) written for spin up and spin down fermions:
\begin{equation}
	H= \int \de x\left\{ \frac{1}{2}\rho_{\uparrow}v_{\uparrow}^{2}
	+\frac{1}{2}\rho_{\downarrow}v_{\downarrow}^{2}
	+\frac{\pi^{2}\hbar^{2}}{6}
    \left(\rho_{\uparrow}^{3}+\rho_{\downarrow}^{3}\right)\right\} \; .
    \label{eq:H_hydro_free_fermions}
\end{equation}

Expanding (\ref{eq:H_hydro_free_fermions}) around the background density $\rho_0=\frac{k_F}{\pi}$ and the background velocity $v_0=0$ up to quadratic terms in $v_{\alpha}$ and $\delta\rho_{\alpha}=\rho_{\alpha}-\rho_{0}$, we obtain the harmonic fluid approximation
\begin{eqnarray}
	H & \approx &  \frac{\rho_{0}}{2}\int \de x\left(v_{\uparrow}^{2}
	+\pi^{2}\hbar^{2}\delta\rho_{\uparrow}^{2}+v_{\downarrow}^{2}
	+\pi^{2}\hbar^{2}\delta\rho_{\downarrow}^{2}\right)
 \nonumber \\
 	& \approx & \frac{\rho_{0}}{4}\hbar^{2}\sum_{\alpha=\uparrow,\downarrow} \int \de x\,
 \left[(\partial_{x}\phi_{R,\alpha})^{2}+(\partial_{x}\phi_{L,\alpha})^{2}\right]
 \label{eq:free_fermion_linearization}
\end{eqnarray}
with right and left bosonic fields defined as $\partial_{x}\phi_{R(L),\alpha}=v_{\alpha}/\hbar\pm \pi\delta\rho_{\alpha}$. This procedure is equivalent to the conventional linear bosonization procedure where the fermionic spectrum is linearized at the Fermi points.

In the spin-charge basis,
\be
  \rho_{c,s} \equiv \rho_{\uparrow}\pm\rho_{\downarrow}
  \qquad {\rm and} \qquad
  v_{c,s}=\frac{v_{\uparrow}\pm v_{\downarrow}}{2} \; ,
  \label{spinchargebasis}
\ee
the harmonic theory (\ref{eq:free_fermion_linearization}) is described
by a sum of two independent harmonic fluid Hamiltonians, one for
charge and the other for spin degrees of freedom
\begin{equation}
	H \approx \frac{\rho_{0}}{4}\int \de x\left(4v_{c}^{2}
	+\pi^{2}\hbar^{2}\delta\rho_{c}^{2}
	+4v_{s}^{2}+\pi^{2}\hbar^{2}\delta\rho_{s}^{2}\right).
 \label{eq:spin-charge-free-fermions}
\end{equation}
After linearization, the quantum Riemann-Hopf Eq. (\ref{FFRiemann}) reduces to
(where $\pm$ stands for $\chi= \{ R,L \}$ respectively)
\begin{equation}
  \dot{k}_{\alpha,\chi} \pm \hbar \pi \rho_0 \; \partial_x
k_{\alpha,\chi} = 0 \; , \qquad
  \alpha = \{ \uparrow, \downarrow \} \; ;
\end{equation}
from which we identify that the quadratic excitations propagate like wave equations with sound velocities $u_{charge}=u_{spin}=\pi\hbar\rho_{0}$, equal for spin and charge.  Turning on interactions between fermions generally renormalizes spin and charge sound velocities differently and results in genuine spin-charge separation at the level of harmonic approximation. The spin-Calogero-Sutherland model happens to be very special in this respect. Despite a non-trivial interaction for spin and charge, their sound velocities remain the same.

Although spin and charge are not truly separated for a free fermion system (and for the sCM), the interaction between spin and charge is absent at the level of harmonic approximation (\ref{eq:spin-charge-free-fermions}). This interaction appears if nonlinear corrections to (\ref{eq:spin-charge-free-fermions}) are taken into account (e.g., by the fully nonlinear Hamiltonian (\ref{eq:H_hydro_free_fermions})) and due to gradient corrections to the hydrodynamics. The latter are not considered in this paper.

In the proper classical limit $\hbar\to 0$ all terms of (\ref{eq:H_hydro_free_fermions}) but the velocity terms vanish (Fermi statistics does not exist for classical particles). Instead, we are interested in a ``semi-classical'' limit in which $\rho\sim v/\hbar$. In this limit we rescale time and velocity by $\hbar$ ($t\to t/\hbar$ and $v\to \hbar v$) and measure everything in length units. This is equivalent to dropping all $\hbar$ from equations. For instance, the Hamiltonian (\ref{eq:H_hydro_free_fermions}) becomes
\begin{equation}
   H=\int \de x\left\{ \frac{1}{2}\rho_{\uparrow}v_{\uparrow}^{2}
   +\frac{1}{2}\rho_{\downarrow}v_{\downarrow}^{2}
   +\frac{\pi^{2}}{6}\left(\rho_{\uparrow}^{3}+\rho_{\downarrow}^{3}\right)\right\}.
 \la{ffhdrop}
\end{equation}
We replace the commutation relations (\ref{rhov-comm}) by the corresponding classical Poisson brackets (for up and down species)
\begin{equation}
   \left\{ \rho_{\alpha}(x), v_{\beta}(y) \right\}
   =\delta_{\alpha\beta}\delta^{\prime}\left(x-y\right) \;
   \label{poisson}
\end{equation}
and consider the classical equations of motion generated by the Hamiltonian together with the Poisson brackets. In the remainder of the paper all hydrodynamic equations are obtained in this {\it semi-classical} limit.

\section{The spin-Calogero model}
 \la{sec:sCM}

In this work we concentrate on the hydrodynamics of sCM (\ref{eq:general_H}) for the case of spin-$1/2$ fermions with an anti-ferromagnetic sign of interaction. It is convenient to impose periodic boundary conditions, i.e. consider particles living on a ring of the length $L$. This Hamiltonian is given by
\begin{equation}
   H=-\frac{\hbar^{2}}{2}\sum_{j=1}^{N}\frac{\partial^{2}}{\partial x_{j}^{2}}
   +\frac{\hbar^{2}}{2}\left(\frac{\pi}{L}\right)^{2}\sum_{j\neq l}
   \frac{\lambda(\lambda-{\bf P}_{jl})}{\sin^{2}\frac{\pi}{L}\left(x_{j}-x_{l}\right)} \;
 \label{eq:h_afm}
\end{equation}
and is known to be integrable\cite{Sutherland-book}.
All eigenstates of (\ref{eq:h_afm}) can be enumerated by the distribution function
\begin{equation}
   \nu(\kappa) =  \nu_{\uparrow}(\kappa) + \nu_{\downarrow} (\kappa).
   \label{eq:nutotal}
\end{equation}
Here, $\kappa$ are integer-valued quantum numbers identifying a given state in a Bethe Ansatz description and $\nu_{\uparrow,\downarrow} (\kappa) = 0,1$ depending on whether a given $\kappa$ is present in the solution of the Bethe Ansatz equations.

The total momentum $P$ and energy $E$ of the eigenstate are given in terms of the distribution function $\nu(\kappa)$ as\cite{sutherland-shastry-1993-PRL,1995-KatoKuramoto}:
\begin{eqnarray}
	P & = & \left(\frac{2\pi}{L}\right) \sum_{\kappa=-\infty}^{+\infty} \kappa \; \nu(\kappa),
 \label{eq:momentum} \\
	E & = & E_{0}+\left(\frac{1}{2}\right)\left(\frac{2\pi}{L}\right)^{2}\epsilon,
 \label{eq:eigenE} \\
	\epsilon &=& \sum_{\kappa=-\infty}^{+\infty}\kappa^{2}\nu(\kappa)
	+\frac{\lambda}{2}\sum_{\kappa,\kappa^{\prime}}|\kappa
	-\kappa^{\prime}|\nu(\kappa)\nu(\kappa^{\prime}),
 \label{eq:eps}
\end{eqnarray}
where $E_{0}=\frac{\pi^{2}\lambda^{2}}{6} N (N^{2}-1)$ is the energy of a reference state\cite{1995-KatoKuramoto}.
The numbers of particles with spin up and spin down are separately conserved in (\ref{eq:h_afm}) and are given by
\be
	N_{\uparrow,\downarrow} = \sum_{\kappa=-\infty}^{+\infty}\nu_{\uparrow,\downarrow}(\kappa).
 \label{eq:N}
\ee
The ground state wave function for (\ref{eq:h_afm}) is\cite{1992-HaHaldane,1995-KatoKuramoto}
\begin{equation}
   	\psi_{GS} =\prod_{j<l} \left| \sin \frac{\pi}{L} (x_{j} - x_{l} ) \right|^{\lambda} \;
   	\prod_{j<l} \left[ \sin \frac{\pi}{L} (x_{j} - x_{l} )
	\right]^{\delta\left(\sigma_{j},\sigma_{l}\right)}
    \exp\left[i\frac{\pi}{2}\mbox{sgn}\left(\sigma_{j}-\sigma_{l}\right)\right]
 \label{eq:gs}
\end{equation}
and corresponds to the distributions \footnote{We neglect $1/N$ corrections and replace combinations like $(N-1)/2$ simply by $N/2$.}
\begin{eqnarray}
   \nu_{\uparrow}(\kappa) & = & \theta(-N_{\uparrow}/2<\kappa<N_{\uparrow}/2) \; ,
   \nonumber \\
   \nu_{\downarrow}(\kappa) & = & \theta(-N_{\downarrow}/2<\kappa<N_{\downarrow}/2) \; .
   \label{eq:nugs}
\end{eqnarray}

\section{Gradientless hydrodynamics of spin-Calogero model}
 \la{sec:hydr}

\begin{figure}
   \dimen0=\textwidth
   \advance\dimen0 by -\columnsep
   \divide\dimen0 by 3
   \noindent\begin{minipage}[t]{\dimen0}
   (a) -- {\bf C}omplete {\bf O}verlap
   \includegraphics[width=\columnwidth]{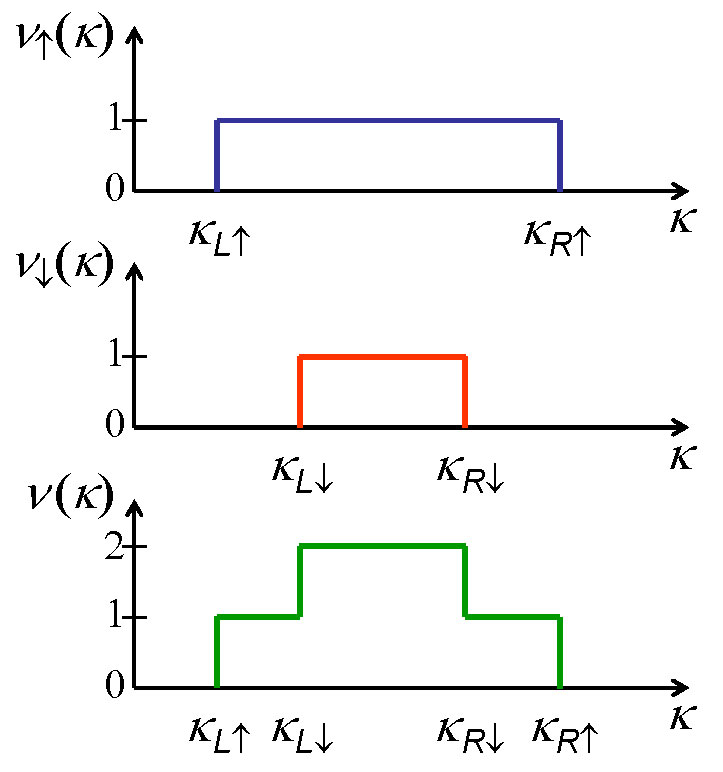}
   \end{minipage}
   \hfill
   \begin{minipage}[t]{\dimen0}
   (b) -- {\bf P}artial {\bf O}verlap
   \includegraphics[width=\columnwidth]{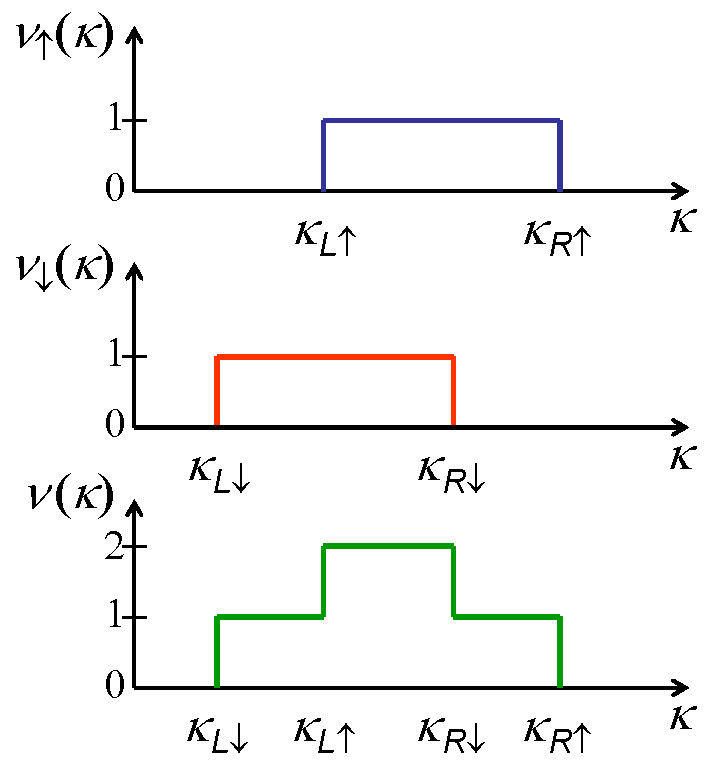}
   \end{minipage}
   \hfill
   \begin{minipage}[t]{\dimen0}
   (c) -- {\bf N}o {\bf O}verlap
   \includegraphics[width=\columnwidth]{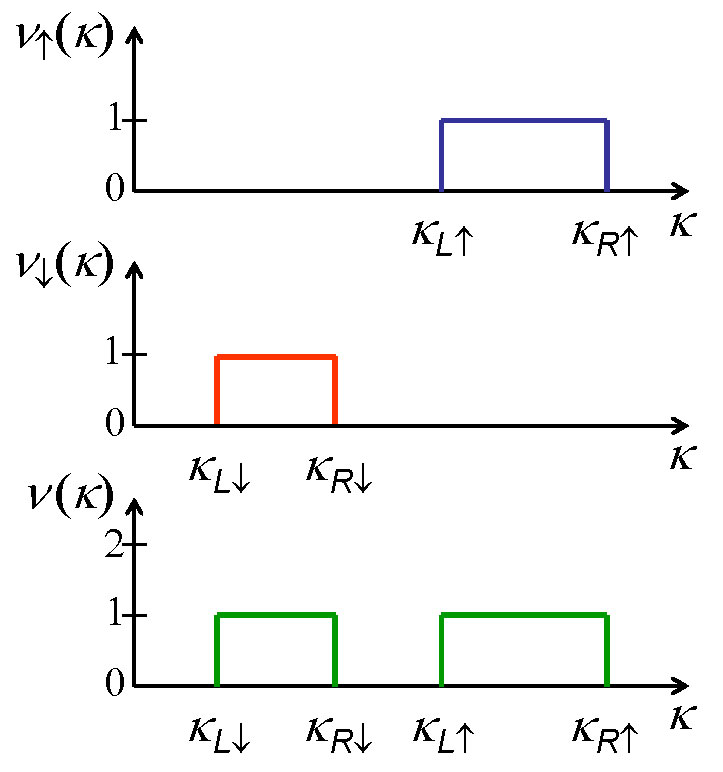}
   \end{minipage}
   \begin{minipage}[t]{\textwidth}
\caption{Distribution functions are shown for the three nonequivalent regimes: {\it Complete Overlap} in (a), {\it Partial Overlap} in (b) and {\it No Overlap} in (c). Three additional regimes exist, but are physically equivalent to the ones considered in these pictures and can be obtained by exchanging ${\bf \uparrow} \leftrightarrow {\bf \downarrow}$.}
   \label{regimes-momentum-distribution}
   \end{minipage}
\end{figure}

Following the example of free fermions, we consider a uniform state specified by the following distributions
\begin{eqnarray}
   \nu_{\uparrow}(\kappa) & = & \theta(\kappa_{L\uparrow}<\kappa<\kappa_{R\uparrow}),
   \label{eq:nu1} \\
   \nu_{\downarrow}(\kappa) & = & \theta(\kappa_{L\downarrow}<\kappa<\kappa_{R\downarrow}).
   \label{eq:nu2}
\end{eqnarray}
This state is the lowest energy state with given numbers of particles, momentum, and total spin current.
It is specified by four integer numbers $\kappa_{L,R;\uparrow,\downarrow}$. All physical quantities such as energy, momentum, and higher integrals of motion of the state can be expressed in terms of these numbers using (\ref{eq:N},\ref{eq:momentum},\ref{eq:eigenE}). These conserved quantities written as integrals over constant quantities are:
\bea
   N_{\alpha} & = & \int \de x \; \rho_{\alpha}
   = \frac{2 \pi}{L} \int \de x \left[ \frac{ \kappa_{R \alpha} - \kappa_{L \alpha}}{2 \pi} \right] \; ,
   \quad \qquad \alpha=\{\uparrow, \downarrow\}
 \\
   P & = & \int \de x \; j_c
   = \left(\frac{2 \pi}{L} \right)^2 \sum_{\alpha=\{\uparrow, \downarrow\}}
   \int \de x \left[\frac{ \kappa_{R \alpha}^2 - \kappa_{L \alpha}^2}{4 \pi} \right] \; .
 \la{mom-int}
\eea
Comparison with (\ref{rhoFF},\ref{JFF}) suggests the following hydrodynamic identifications:
\bea
   v_{\uparrow} \pm \pi \rho_{\uparrow} & \equiv &
   \frac{2 \pi}{L} \; \kappa_{(R,L);\uparrow} \; ,
   \label{eq:kup} \\
   v_{\downarrow} \pm \pi \rho_{\downarrow} & \equiv &
   \frac{2 \pi}{L} \; \kappa_{(R,L);\downarrow} \; .
   \label{eq:k3kdown}
\eea
In the main body of the paper we use $v_{\uparrow,\downarrow}$ and refer to them as to  ``velocities''. At this point they have been introduced ``by analogy'' with the case of free fermions.  In  Appendices \ref{app:AABsCM},\ref{app:velocities},\ref{app:allcases} we show that these velocities are indeed conjugated to the corresponding densities and explain their relations to the true hydrodynamic velocities. In fact, in the most interesting case to us, namely the CO regime (see below) these velocities coincide with the true hydrodynamic velocities defined in Appendix \ref{app:velocities}.
The total momentum (\ref{mom-int}) of the system in terms of (\ref{eq:kup},\ref{eq:k3kdown}) is
\begin{equation}
   P = \int \de x \big( \rho_{\uparrow} v_{\uparrow}
   + \rho_{\downarrow} v_{\downarrow} \big) \; .
   \label{eq:momentum_g}
\end{equation}
One can also express the energy (\ref{eq:eigenE}) in terms of these hydrodynamic variables. Because of the non-analyticity (presence of an absolute value) in formula (\ref{eq:eigenE}) it is convenient to consider different physical regimes.  These regimes are defined by the mutual arrangement of the supports of the distribution functions (\ref{eq:nu1},\ref{eq:nu2}). There are six different regimes, that reduce to three physically non-equivalent ones using the permutation ${\bf \uparrow} \leftrightarrow {\bf \downarrow}$. The distributions corresponding to different regimes are shown in   Fig.~\ref{regimes-momentum-distribution}:
\begin{itemize}
	\item{\textit{Complete Overlap (CO) regime}}. The support of $\nu_\downarrow$ is completely contained in $\nu_\uparrow$ (or vice versa). This is the regime considered in Ref. \cite{1992-HaHaldane}, where its exact solution was given.
    \item{\textit{Partial Overlap (PO) regime}}. The supports of $\nu_\uparrow$ and of $\nu_\downarrow$ only partially overlap.
    \item{\textit{No Overlap (NO) regime}}. The supports of $\nu_\uparrow$ and of $\nu_\downarrow$ do not overlap at all.
\end{itemize}
Notice that the small fluctuations around the singlet ground state (with $\rho_{s}=0$) belong to first two regimes.

\begin{figure}
   \includegraphics[width=10cm]{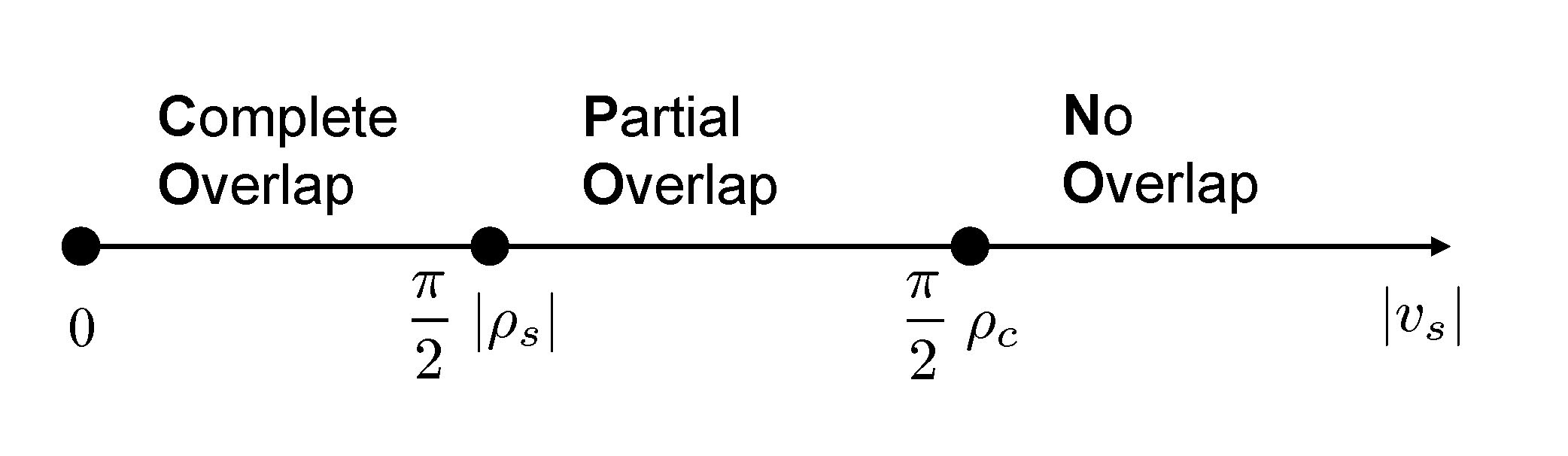}
   \caption{\label{fig:Diagram-capturing-all}Diagram capturing all cases}
\end{figure}

In terms of the hydrodynamic variables the three regimes are summarized in Fig.~\ref{fig:Diagram-capturing-all} and are defined by the following inequalities:
\begin{eqnarray}
   \mbox{Complete Overlap} & \rightarrow & |v_{s}|<\frac{\pi}{2} \; |\rho_{s}| \; ,
   \label{eq:COR} \\
   \mbox{Partial Overlap} & \rightarrow & \frac{\pi}{2} \; |\rho_{s}| < |v_{s}| < \frac{\pi}{2} \; \rho_{c} \; ,
   \label{eq:inter} \\
   \mbox{No Overlap} & \rightarrow & \frac{\pi}{2} \; \rho_{c} <  |v_{s}| \; ,
   \label{eq:lfr}
\end{eqnarray}
where we switched to the spin and charge degrees of freedom defined by (\ref{spinchargebasis}).
To simplify the presentation we give here formulas only for the CO regime
\begin{equation}
   -\frac{\pi\rho_{s}}{2}<v_{s}<\frac{\pi\rho_{s}}{2} \; ,
   \label{eq:Haldnae1inequality}
\end{equation}
where we also assumed that $\rho_{s}>0$. The opposite case $\rho_{s}<0$ can be obtained exchanging up and down variables. The other regimes and formulae valid for all regimes are considered  in detail in Appendix \ref{app:allcases}.

In the CO regime (\ref{eq:Haldnae1inequality}), the Hamiltonian can be written as
\begin{eqnarray}
   H_{\textrm{CO}} & = & \int \de x \; \Biggl\{
   \frac{1}{2} \rho_{\uparrow} v_{\uparrow}^{2}
   + \frac{1}{2} \rho_{\downarrow} v_{\downarrow}^{2}
   + \frac{\lambda}{2} \rho_{\downarrow}
   \big( v_{\uparrow} - v_{\downarrow} \big)^{2}
   \nonumber \\
   && \qquad \quad + \frac{\pi^{2}\lambda^{2}}{6} \rho_{c}^{3} +
   \frac{\pi^{2}}{6} \big(\rho_{\uparrow}^{3} + \rho_{\downarrow}^{3} \big) + \frac{\lambda\pi^{2}}{6} \big(2\rho_{\uparrow}^{3}
   + 3\rho_{\uparrow}^{2} \rho_{\downarrow}
   + 3\rho_{\downarrow}^{3} \big) \Biggr\} \; .
   \label{eq:CO_hamiltonian}
\end{eqnarray}
It is obtained by expressing (\ref{eq:eigenE},\ref{eq:eps}) in terms of hydrodynamic variables (\ref{eq:kup},\ref{eq:k3kdown}) using (\ref{eq:Haldnae1inequality}). As in the case of free fermions (see Sec.~\ref{sec:freeferm}) we now consider $\rho_{\uparrow,\downarrow}(x,t)$ and $v_{\uparrow,\downarrow}(x,t)$ as space and time dependent classical hydrodynamic fields with Poisson brackets (\ref{poisson}). Of course, going from the energy of the uniform state (\ref{eq:nu1},\ref{eq:nu2}) to the nonuniform hydrodynamic state we neglected gradients of density and velocity fields. We refer to this approximation as to \textit{gradientless hydrodynamics}. The equations of motion generated by the Hamiltonian (\ref{eq:CO_hamiltonian}) with Poisson brackets (\ref{poisson}) can be used only when gradients can be neglected compared to the gradientless terms. This means that one can use this gradientless hydrodynamics only at relatively small times (compared to the time of the gradient catastrophe, see the discussion below).

Before analyzing more general case let us consider some special limits of
(\ref{eq:CO_hamiltonian}).

\subsubsection{Spinless limit}

In the fully polarized state $\rho_{\downarrow}=0$ we obtain
from (\ref{eq:CO_hamiltonian}) the  gradientless Hamiltonian
for spinless Calogero-Sutherland model
\be
	H^{\mbox{spinless}}=\int_{-\infty}^{+\infty}dx
	\left\{ \frac{1}{2}\rho v^{2}+\frac{\pi^{2}}{6}\left(\lambda+1\right)^{2}\rho^{3}\right\},
 \label{spingless}
\ee
where we dropped  the subscript $\uparrow$. The hydrodynamics (\ref{spingless}) was used in \cite{2005-Abanov-LesHouches} to calculate the leading term of an asymptotics of a particular correlation function (Emptiness Formation Probability) for the Calogero-Sutherland model.  It  can be, of course, obtained by dropping gradient terms in the exact hydrodynamics derived using collective field theory \cite{1983-AndricJevickiLevine,1988-AndricBardek,1995-Polychronakos}.

\subsubsection{\texorpdfstring{$\lambda=0$}{} -- free fermions with spin}

At the particular value $\lambda=0$ the sCM reduces to free fermions with spin and the Hamiltonian (\ref{eq:CO_hamiltonian}) becomes the collective Hamiltonian for free fermions (\ref{ffhdrop}).

\subsubsection{\label{sub:-large-lambda}\texorpdfstring{$\lambda\rightarrow\infty$}{} limit.}

In the limit of large coupling constant $\lambda\to\infty$ the particles form a rigid lattice and charge degrees of freedom essentially get frozen \cite{1993-Polychronakos}. We expect to arrive to an effective spin dynamics equivalent to the Haldane-Shastry model \cite{1988-Haldane-HS,1988-Shastry-HS} (see Appendix \ref{app:hdHSM}). This reduction to the Haldane-Shastry model is usually referred to as \textit{freezing trick} \cite{1993-Polychronakos}. We analyze this reduction in more detail in Sec.~\ref{sec:Lattice}.

\section{Equations of motion and separation of variables}
 \la{sec:eqm}

\subsection{Equations of motion}

The classical gradientless hydrodynamics for sCM is given by the Hamiltonian (\ref{eq:CO_hamiltonian}) with canonic Poisson's brackets (\ref{poisson}).
The classical evolution equations generated by this Hamiltonian are
\begin{eqnarray}
	\dot{\rho}_{\uparrow} & = & -\partial_{x}\left\{ \rho_{\uparrow}v_{\uparrow}
	+\lambda\rho_{\downarrow}\left(v_{\uparrow}-v_{\downarrow}\right)\right\},
 \nonumber \\
	\dot{\rho}_{\downarrow} & = & -\partial_{x}\left\{ \rho_{\downarrow}v_{\downarrow}
	-\lambda\rho_{\downarrow}\left(v_{\uparrow}-v_{\downarrow}\right)\right\},
 \nonumber \\
	\dot{v}_{\uparrow} & = & -\partial_{x}\Biggl\{\frac{v_{\uparrow}^{2}}{2}
	 +\frac{\pi^{2}\lambda^{2}}{2}\left(\rho_{\uparrow}+\rho_{\downarrow}\right)^{2}
	 +\lambda\pi^{2}\left(\rho_{\uparrow}^{2}+\rho_{\uparrow}\rho_{\downarrow}\right)
	+\frac{\pi^{2}}{2}\rho_{\uparrow}^{2}\Biggr\},
 \label{updownhydro} \\
	\dot{v}_{\downarrow} & = & -\partial_{x}\Biggl\{\frac{v_{\downarrow}^{2}}{2}
	+\frac{\lambda}{2}\left(v_{\uparrow}-v_{\downarrow}\right)^{2}
	 +\frac{\pi^{2}\lambda^{2}}{2}\left(\rho_{\uparrow}+\rho_{\downarrow}\right)^{2}
	 +\frac{\lambda\pi^{2}}{2}\left(\rho_{\uparrow}^{2}+3\rho_{\downarrow}^{2}\right)
	+\frac{\pi^{2}}{2}\rho_{\downarrow}^{2}\Biggr\}.
 \nonumber
\end{eqnarray}
This is the system of continuity and Euler's equations for two coupled fluids (with spin up and spin down). We can also rewrite it in terms of spin and charge variables (\ref{spinchargebasis})
\begin{eqnarray}
	\dot{\rho}_{c} & = & -\partial_{x}\left\{\rho_{c}v_{c}+\rho_{s}v_{s}\right\},
 \nonumber \\
	\dot{\rho}_{s} & = & -\partial_{x}\left\{\rho_{s}(v_{c}-2\lambda v_{s})
	+(2\lambda+1)\rho_{c}v_{s}\right\},
 \nonumber \\
	\dot{v}_{c} & = &  -\partial_{x}\left\{\frac{v_{c}^{2}}{2}+(2\lambda+1)\frac{v_{s}^{2}}{2}
	 +\frac{\pi^{2}}{8}\left[(2\lambda+1)^{2}\rho_{c}^{2}+(2\lambda+1)\rho_{s}^{2}\right]\right\},
 \la{cshydro} \\
	\dot{v}_{s} & = & -\partial_{x}\left\{v_{c}v_{s}-\lambda v_{s}^{2}
	 +\frac{\pi^{2}}{4}\rho_{s}\left[(2\lambda+1)\rho_{c}-\lambda\rho_{s}\right]\right\}.
 \nonumber
\end{eqnarray}
One can see that spin and charge are not decoupled. It turns out, however,  that the variables nevertheless separate and the system of four coupled equations (\ref{updownhydro}) can be written as four decoupled Riemann-Hopf equations (similar to (\ref{FFRiemann})) for a special linear combinations of density and velocity fields. In the following we study the interaction
of spin and charge governed by the above equations.

\subsection{Free fermions (\texorpdfstring{$\lambda=0$}{}) and Riemann-Hopf equation}
\label{ffRE}

At $\lambda=0$  equations (\ref{updownhydro}) become the hydrodynamic equations for free fermions. Fluids corresponding to up and down spin are completely decoupled
\begin{eqnarray}
	\dot{\rho}_{\uparrow,\downarrow}
	& = & -\partial_{x}\left\{ \rho_{\uparrow,\downarrow}v_{\uparrow,\downarrow}\right\},
 \label{eq:rdorfree} \\
	\dot{v}_{\uparrow,\downarrow}
	& = & -\partial_{x}\left\{ \frac{1}{2}v_{\uparrow,\downarrow}^{2}
	+\frac{\pi^{2}}{2}\rho_{\uparrow,\downarrow}^{2}\right\}.
 \label{eq:vdotfree}
\end{eqnarray}
Let us introduce the following linear combinations of densities and velocities
\begin{eqnarray}
	k_{R\uparrow,L\uparrow} & = & v_{\uparrow} \pm \pi\rho_{\uparrow},
 \nonumber \\
	k_{R\downarrow,L\downarrow} & = & v_{\downarrow} \pm \pi\rho_{\downarrow}.
 \la{uff}
\end{eqnarray}
These combinations are nothing else but right and left Fermi momenta of free fermions. All of them satisfy the so-called Riemann-Hopf equation
\be
	u_{t}+uu_{x}=0.
 \la{RE}
\ee
The equation is the same for all four combinations $u=k_{R,L;\uparrow,\downarrow}$ and the system (\ref{eq:rdorfree},\ref{eq:vdotfree}) is equivalent to four decoupled Riemann-Hopf equations.

The Riemann-Hopf equation (\ref{RE}) is easily solvable with the general solution given implicitly by
\be
	u = u_{0}(x-ut).
 \la{REsol}
\ee
Here $u_{0}(x)$ is an initial profile of $u(x,t)$ at $t=0$. One should solve (\ref{REsol}) with respect to $u$ and find $u(x,t)$ - the solution of (\ref{RE}) with $u(x,t=0)=u_{0}(x)$. The solution (\ref{REsol}) can also be written in a parametric form
\bea
	x &=& y +t\, u_{0}(y),
 \nonumber \\
 	u(x,t) &=& u_{0}(y).
 \la{REsol1}
\eea
This solution corresponds to the ``Lagrangian picture'' of fluid dynamics and states that points in the $x-u$ plane are just translated along $x$ with velocity $u$, i.e., $(x,u_{0})\to (x+t\,u_{0},u_{0})$. This picture is especially useful to solve (\ref{RE}) numerically.

We notice here that the nonlinear dynamics (\ref{RE}) without dispersive (higher gradient) terms is ill defined at large times. For any initial profile $u_{0}(x)$, at large times $t>t_{c}$ infinite gradients $u_{x}$ will develop - \textit{gradient catastrophe} - and solutions of (\ref{REsol}) will become multi-valued. The classical equation (\ref{RE}) will not have a meaning for $t>t_{c}$. We refer to the time $t_{c}$ (function of the initial profile) as to the gradient catastrophe time. The gradientless hydrodynamics is applicable only for times smaller that $t_{c}$. \footnote{We notice here that for a free fermion system it is possible to assign the meaning even to the multi-valued solution of (\ref{REsol}) for $t>t_{c}$. It is the boundary of the support of the Wigner distribution in the one-particle phase space. In this paper we restrict ourselves to times less than the time of gradient catastrophe and assume that (\ref{RE}) has a well-defined single-valued solution.} We will discuss in more detail about validity of gradientless hydrodynamics in Sec.~\ref{sec:pictures}.

\begin{figure}
	\includegraphics[width=0.45\columnwidth]{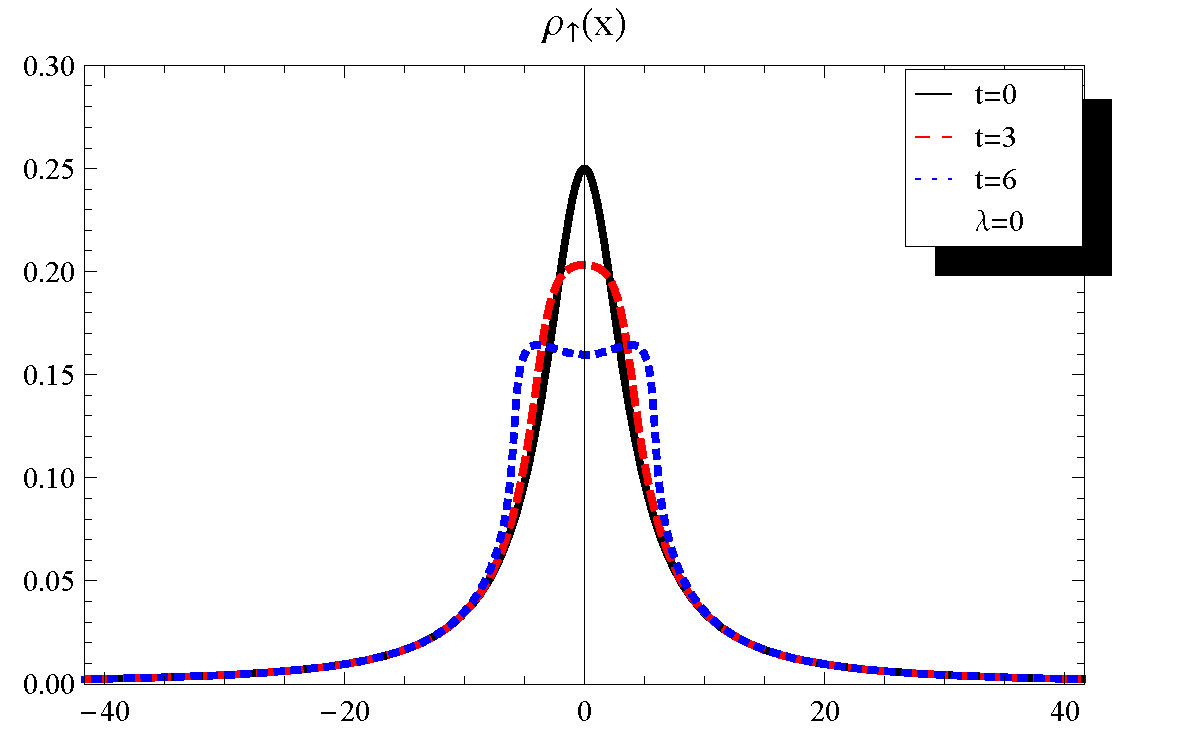}
	\includegraphics[width=0.45\columnwidth]{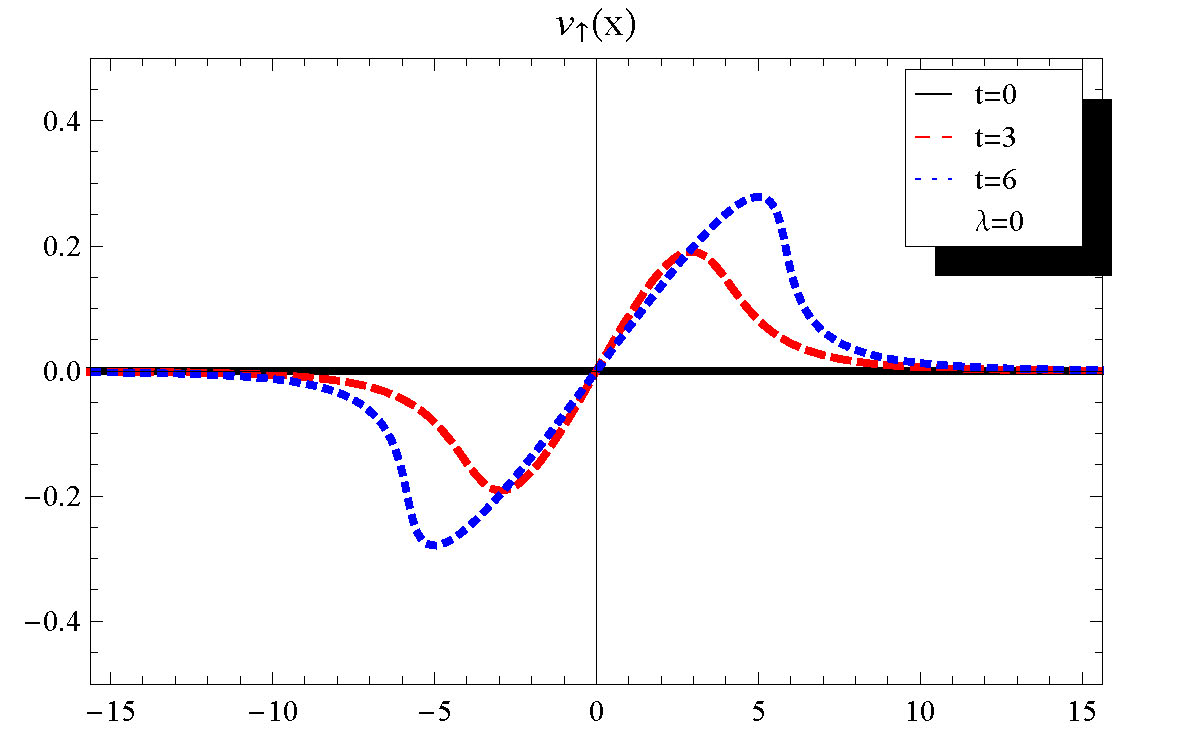}
	\caption{
		\label{fig:Dynamics-of-density}
	Dynamics of density field $\rho_{\uparrow}(x)$ (left panel)
	and of velocity field $v_{\uparrow}(x)$ (right panel) for free a fermion case ($\lambda=0$).
	The initial density profile at $t=0$ is a Lorentzian (\ref{Lorentzian}) of height $h=0.25$ and
	half-width $a=4$. The initial velocity is zero.}
\end{figure}

We present a simple illustration of the density and velocity dynamics for free fermion system in Fig.~\ref{fig:Dynamics-of-density}. It is enough to consider only up-spin as the evolution of up and down spins is decoupled. We chose the initial profile of the density as Lorentzian with the half-width $a$ and height $h$
\be
	\rho_{0\uparrow}(x)=\frac{h}{1+(x/a)^{2}}
 \la{Lorentzian}
\ee
and an initial velocity zero. We find the initial profiles of $k_{\uparrow;R,L}$ using (\ref{uff}). Then we solve the Riemann-Hopf equations (\ref{RE}) using (\ref{REsol1}) and find the density and velocity at any time inverting (\ref{uff}). We remark that for an arbitrary smooth bump of height $h$ and width $a$ the gradient catastrophe time can be estimated as $t_{c}\approx\frac{a}{h}$.
For the evolution given by (\ref{RE}) with an initial Lorentzian profile ($u_{0}(x)$ given by (\ref{Lorentzian})) one can compute the gradient catastrophe time exactly. An infinite gradient $\partial_{x}u\to \infty$ develops at the time
\be
	t_{c}=\frac{8}{3\sqrt{3}}\frac{a}{h}.
 \la{catastrphetimelor}
\ee
For arbitrary initial conditions we compute the gradient catastrophe time numerically.

\subsection{Riemann-Hopf Equations for sCM}

Although the system of equations (\ref{updownhydro}) is a system of four coupled nonlinear equations, it allows for a separation of variables. Introducing the linear combinations of fields
\begin{eqnarray}
	k_{R\uparrow,L\uparrow}
	& = & v_{\uparrow} \pm \pi\left[(\lambda+1)\rho_{\uparrow}+\lambda\rho_{\downarrow}\right]
	=  \left(v_{\uparrow}\pm\pi\rho_{\uparrow}\right) \pm \lambda \pi\rho_{c},
 \label{eq:uup}\\
	k_{R\downarrow,L\downarrow}
	& = & (\lambda+1)v_{\downarrow}-\lambda v_{\uparrow} \pm \pi(2\lambda+1)\rho_{\downarrow}
	= \left(v_{\downarrow}\pm\pi\rho_{\downarrow}\right)
	+\lambda(-2v_{s}\pm 2\pi \rho_{\downarrow})
 \label{eq:udown}
\end{eqnarray}
we obtain the Riemann-Hopf equation (\ref{RE}) separately for all four $u=k_{L,R,\uparrow,\downarrow}$.

This property of variable separation is shared with the free fermion case Sec.\ref{ffRE}. We notice, however, that in the case of sCM, variables separate only in gradientless approximation. The gradient terms neglected in this paper will couple the hydrodynamic equations in an essential non-separable\footnote{At least one will not be able to separate variables considering simple linear combinations of fields. To the best of our knowledge variables in sCM do not separate or, at least, an appropriate change of variables has not been found yet.} way.

The separation of variables in terms of (\ref{eq:uup},\ref{eq:udown}) is not so surprising. One can recognize (\ref{eq:uup},\ref{eq:udown}) as dressed (physical) ``Fermi'' momenta of (asymptotic) Bethe Ansatz.  The integrals of motion of sCM are separated in terms of these Fermi momenta and the same is true for the equations of motion. We do not interrupt the presentation with this connection with the Bethe Ansatz solution of sCM but devote the Appendix \ref{app:AABsCM} to this purpose.

\begin{figure}
   \includegraphics[width=10cm]{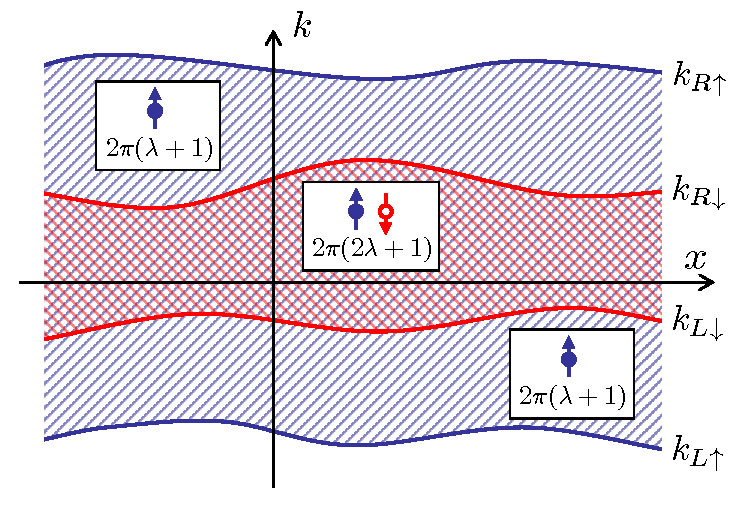}
   \caption{\label{fig:phase-space} Phase-space diagram of a hydrodynamic state characterized by four space-dependent Fermi momenta.}
\end{figure}

It is convenient to summarize the gradientless hydrodynamics of sCM by the picture in a ``single-particle'' phase space showing space-dependent Fermi momenta. \footnote{We would like to thank A. Polychronakos who encouraged us to present this picture.} We plot the space-dependent Fermi momenta in an $x-k$ plane as four smooth lines. In the CO regime considered here (see Appendix \ref{app:allcases} for other regimes) the Fermi momenta are ordered as
\be
	k_{L\uparrow}(x) < k_{L\downarrow}(x) < k_{R\downarrow}(x) < k_{R\uparrow}(x).
 \la{COdmomenta}
\ee
We fill the space between those lines with particles obeying the following rules of particles with fractional exclusion statistics \cite{1997-KatoYamamotoArikawa} (see Appendix \ref{app:AABsCM})
(i) each particle occupies a phase space volume $2\pi(\lambda+1)$ if there are no particles of the other species in this volume, (ii) two particles with opposite spins occupy a phase space volume $2\pi (2\lambda+1)$ (or $2\pi (\lambda+1/2)$ per particle). The velocity $v_{\uparrow}(x)$ is visualized as a center of a spin-up stripe on Figure \ref{fig:phase-space} (see (\ref{vupdown})). The interpretation of $v_{\downarrow}(x)$ is a bit less straightforward. It should be thought as a weighted average of positions of centers of both stripes (\ref{vupdown}).

\section{Freezing trick and hydrodynamics of Haldane-Shastry model}
 \la{sec:Lattice}

Here we consider the limit of large coupling constant $\lambda\to \infty$. In this limit we expect that particles form a one-dimensional lattice and only spin dynamics is important at low energies. We refer to this limit as to a freezing of the charge. We are interested in fluctuations around the uniform state with a given charge density. It can be seen from Fig.~\ref{fig:phase-space} that particles occupy the volume $2\pi(\lambda+1/2)$ of the phase space when both species are present. Therefore, the natural expansion parameter is $\mu=\lambda+1/2$ instead of $\lambda$.\footnote{Of course, first orders of the expansion are not sensitive to this shift.} We will see that the leading in $\mu$ term of the dynamics results in charge freezing, while the next to leading term gives the non-trivial spin dynamics of the lattice model known as the Haldane-Shastry model \cite{1988-Haldane-HS,1988-Shastry-HS}
\begin{equation}
	H_{\rm HSM}=2\sum_{j<l}\frac{S_{j}\cdot S_{l}}{\left(j - l \right)^{2}}.
 \label{eq:Haldane_shastry}
\end{equation}
This model is known to be integrable.\cite{Sutherland-book} The freezing procedure described is referred to as ``freezing trick'' and was introduced by Polychronakos \cite{1993-Polychronakos}. Our goal is to implement the procedure in a hydrodynamic description.

Before proceeding to a regular expansion of the equations of motion we start with a heuristic argument. We rewrite the hydrodynamic Hamiltonian (\ref{eq:CO_hamiltonian}) in terms of spin and charge variables (\ref{spinchargebasis}) and consider first the two leading terms in a $1/\mu$ expansion
\bea
	H &=& \int \de x\, \left\{\frac{1}{2}\rho_{c}v_{c}^{2}+\rho_{s}v_{c}v_{s}+\mu\rho_{c}v_{s}^{2}-\left(\mu-\frac{1}{2}\right)\rho_{s}v_{s}^{2}
	+\frac{\pi^{2}\mu^{2}}{6}\rho_{c}^{3}
	+\frac{\pi^{2}}{4}\mu\rho_{c}\rho_{s}^{2}
	-\frac{\pi^{2}}{12}\left(\mu-\frac{1}{2}\right)\rho_{s}^{3}\right\}
\label{mu-rep} \\
	& = & \int \de x\, \left\{
	\frac{\pi^{2}}{6}\mu^{2}\rho_{c}^{3}
	+\mu\Biggl[\rho_{c}v_{s}^{2}-\rho_{s}v_{s}^{2}
	+\frac{\pi^{2}\rho_{c}\rho_{s}^{2}}{4}
	- \frac{\pi^{2}\rho_{s}^{3}}{12}\Biggr]
	+O(1)\right\}.
 \label{eq:CO-case-freezing}
\eea
The first term proportional to $\mu^{2}$ comes from the energy of a static lattice while the second term proportional to $\mu$ gives the Hamiltonian of the Haldane-Shastry model in the hydrodynamic formulation (see Appendix \ref{app:hdHSM}), i.e., describes the spin dynamics. Note that $\rho_{c}$ here should be considered as a constant equal to the inverse lattice spacing of the charge lattice.

To build a systematic expansion in $1/\mu$ we go to the hydrodynamic evolution equations given in (\ref{cshydro}).
We introduce the following series in $1/\mu=1/(\lambda+1/2)$ for the space-time dependent fields.
\begin{eqnarray*}
	u & = & u^{(0)}+\frac{1}{\mu}u^{(1)}
	+\frac{1}{\mu^{2}}u^{(2)}+...
 \\
	u & \to  & \rho_{c}, v_{c}, \rho_{s}, v_{s}
\end{eqnarray*}
and re-scale time $t = \tau/\mu$ (or $\partial_{t} = \mu\partial_{\tau}$).
We substitute these expansions into (\ref{cshydro}) and
compare order by order in $\mu$. Let us consider few leading orders explicitly.

\subsection{\texorpdfstring{$O(\mu)$}{}}

In this order the only non-trivial equation gives
\begin{equation}
	0=-\partial_{x}\left[\rho_{c}^{(0)^{2}}\right]
 \label{eq:ol1}
\end{equation}
and implies that $\rho_{c}^{(0)}$ is constant in space.

\subsection{\texorpdfstring{$O(1)$}{}}

At this order we have
\begin{eqnarray}
	\dot{\rho}_{c}^{(0)} & = & 0,
 \label{eq:o1rhoc} \\
	\dot{\rho}_{s}^{(0)} & = & -\partial_{x}\left\{ 2\rho_{c}^{(0)} v_{s}^{(0)}
	-2 \rho_{s}^{(0)} v_{s}^{(0)}\right\},
 \label{eq:o1rhos}\\
	\dot{v}_{c}^{(0)} & = & -\partial_{x}\left\{ v_{s}^{(0)^{2}}
	+\pi^{2}\rho_{c}^{(0)}\rho_{c}^{(1)}
	+   \frac{\pi^{2}}{4}\rho_{s}^{(0)^{2}}\right\},
 \label{eq:o1vc}\\
	\dot{v}_{s}^{(0)} & = & -\partial_{x}\left\{ -v_{s}^{(0)^{2}}
	+\frac{\pi^{2}}{2}\rho_{c}^{(0)}\rho_{s}^{(0)}
	-\frac{\pi^{2}}{4}\rho_{s}^{(0)^{2}}\right\}.
 \label{eq:o1vs}
\end{eqnarray}
Combining  (\ref{eq:ol1}) and (\ref{eq:o1rhoc}) we see that $\rho_{c}^{(0)}$ is a constant independent of space-time. The evolution equations (\ref{eq:o1rhos}) for spin density, $\dot{\rho}_{s}^{(0)}$ and (\ref{eq:o1vs}) for the spin
velocity $\dot{v}_{s}^{(0)}$ do not depend on the dynamics of the charge
and are precisely the ones obtained for the Haldane-Shastry model (compare to (\ref{HShydro})).
We refer the reader to the Appendix \ref{app:hdHSM} for more details on the hydrodynamics of the Haldane-Shastry model.

Equation (\ref{eq:o1vc}) is important in resolving a well-known ``paradox''. In the original spin-Calogero model the momentum of the system is identical to the total charge current since all particles in the model have the same charge. On the other hand in the Haldane-Shastry model the momentum is carried by spin excitations and superficially no charge motion is involved. One can ask how this is compatible with getting the Haldane-Shastry model in the limit $\lambda\to \infty$ from the spin Calogero model. Equation (\ref{eq:o1vc}) is necessary to make sure that the the current density $j(x)=\rho_{c}v_{c}+\rho_{s}v_{s}$ is globally conserved at a given
order in $1/\mu$. Since  $\rho_{c}^{(0)}$ is a constant in space-time we expect from (\ref{eq:o1rhos}) and (\ref{eq:o1vs}) that $v_{c}^{(0)}$ evolves according to (\ref{eq:o1vc}) to ensure that the current density is conserved. As a result, there is a charge motion associated with the momentum but in the large $\lambda$ limit this ``recoil'' momentum is absorbed by the whole charge lattice.

\subsection{\texorpdfstring{$O(1/\mu)$}{}}

For the sake of brevity we do not write down the equations at this
order but make some comments instead. In the previous order,
$O(1)$ we noticed (see eqs. (\ref{eq:o1rhos}) and (\ref{eq:o1vs}) that
spin degrees of freedom evolve as the charge is essentially frozen and at that order there is no feedback of the charge degrees of freedom on the spin. However, in the order $O(1/\mu)$ we have feedback terms in both
evolution equations for $\rho_{s}$ and $v_{s}$. As an example we have $\dot{\rho}_{s}^{(1)}=-\partial_{x}\left\{ ...+2v_{s}^{(0)}\rho_{c}^{(1)}+v_{c}^{(0)}\rho_{s}^{(0)}+...\right\} $
and $\dot{v}_{s}^{(1)}=-\partial_{x}\left\{ ...+v_{c}^{(0)}v_{s}^{(0)}+\frac{\pi^{2}}{2}\rho_{c}^{(1)}\rho_{s}^{(0)}+....\right\} $
which clearly show that there is a charge feedback into the spin sector.

\subsection{Evolution equations for Haldane-Shastry model from the freezing trick}

The shortest way to evolution equations for Haldane-Shastry model is to take $\lambda\to \infty$ limit directly in Riemann-Hopf equations (\ref{RE}).  After rescaling time $t=\tau/\mu$ we have
\be
	\tilde{k}_{\tau} +\tilde{k}\tilde{k}_{x} =0,
 \la{tildeRH}
\ee
where $\tilde{k}=k/\mu = k/(\lambda+1/2)$. In the large $\lambda$ limit we have using (\ref{eq:uup},\ref{eq:udown}) $\tilde{k}_{R\uparrow,L\uparrow}\to \pm \pi\rho_{c}$ and $\tilde{k}_{R\downarrow,L\downarrow}=-2v_{s}\pm 2\pi \rho_{\downarrow}$. Then the equation (\ref{tildeRH}) gives evolution equations for Haldane-Shastry model with (\ref{HSMident},\ref{frid}).

\section{Illustrations}
 \la{sec:pictures}

It is relatively simple to obtain the evolution of arbitrary (smooth) initial density and velocity profiles solving equations of the gradientless hydrodynamics (\ref{cshydro}) numerically. One can do it very effectively using the fact that the dynamics is separated into four Riemann-Hopf equations (\ref{RE}) and using their general solutions (\ref{REsol1}). In this section we give numerical results for charge and spin dynamics  corresponding to a relaxation of a (spin) polarized center. These results show that due to nonlinearity of equations spin can drag charge in spin-Calogero model. We notice here that in the examples considered in this section the dynamics belongs to CO regime. \footnote{In exotic cases involving boundaries between CO and PO regimes one notices singularities developing at the boundary and we expect gradient corrections to correct these singularities.}

\subsection{Charge dynamics in a spin-singlet sector}

As a first example we consider initial conditions $\rho_{s},v_{s}=0$ and some arbitrary initial conditions for $\rho_{c}$ and $v_{c}$. It is easy to see from (\ref{cshydro}) that spin density and spin velocity remain zero at any time while charge degrees of freedom satisfy
\bea
	\dot\rho_{c} &=& -\partial_{x}(\rho_{c}v_{c}),
 \nonumber \\
 	\dot{v}_{c} &=& -\partial_{x}\left\{\frac{v_{c}^{2}}{2}+\frac{\pi^{2}\left(\lambda+\frac{1}{2}\right)^{2}\rho_{c}^{2}}{2}\right\}.
 \label{spinsinglet}
\eea
Hydrodynamics (\ref{spinsinglet}) is identical to the one of the Calogero-Sutherland model with one species of particles (except for the change $\lambda+1\to \lambda+1/2$). It can be written as a system of two Riemann-Hopf equations (\ref{RE}) for fields $v_{c}\pm \pi(\lambda+1/2)\rho_{c}$.

We conclude that the charge dynamics does not affect spin in a spin-singlet state at least in the gradientless limit.  It is interesting to see how spin dynamics affects the charge one.

\subsection{Dynamics of a polarized center}

To see how spin drags charge we start with an initial configuration with static and uniform charge background. We assume that initially there is no spin current  but there is a non-zero polarization given by a Lorentzian profile:
\bea
 	t=0: \qquad\qquad
	\rho_{c}=1, \qquad v_{c}=0, \qquad v_{s}=0, \qquad
	\rho_{s} = \frac{h}{1+(x/a)^{2}},
 \la{initLor}
\eea
i.e., there is an excess of particles with spin up over particles with spin down near the origin. The maximal polarization is $h$ and a half-width of the polarized center is $a$. As an illustration of spin and charge dynamics we present a solution of (\ref{cshydro}) with initial conditions (\ref{initLor}) corresponding to $h=0.25$ and $a=4$. Some important comments are in order.

\subsubsection{Applicability of gradientless hydrodynamics}

The hydrodynamic equations we use (\ref{cshydro}) neglect gradient corrections and, therefore, are approximate. They can be applied only under the condition that the neglected higher gradient terms are small compared to the terms taken into account in (\ref{cshydro}). Of course, the exact criteria can be written only when the form of the higher gradient terms are known explicitly. Here, we are going to use a much simpler criterion. We require that all fields change slowly at the scale of the inter-particle spacing. The uniform background $\rho_{c}=1$ defines the inter-particle spacing and the characteristic scale for hydrodynamic fields to be $1$ and we require $\partial_{x}f\ll 1$ for all fields at all $x$ and $t$ that we consider.

One can easily check that $\partial_{x}\rho_{s}(x,t=0) \ll 0.1$  for all $x$ with the initial profile (\ref{initLor}) (in fact, the maximal derivative is approximately $0.041$). Because of the gradient catastrophe this condition will be broken at some time and we can trust the results obtained from (\ref{cshydro}) only up to that time. To be well within this criterion all our fields satisfy $\partial_{x}f<0.3$ at any given time.

Let us start with the solutions for the case of free fermions, i.e., $\lambda=0$.

\subsubsection{Free fermions with spin: \texorpdfstring{$\lambda=0$}{}}

\begin{figure}
\includegraphics[width=0.45\columnwidth]{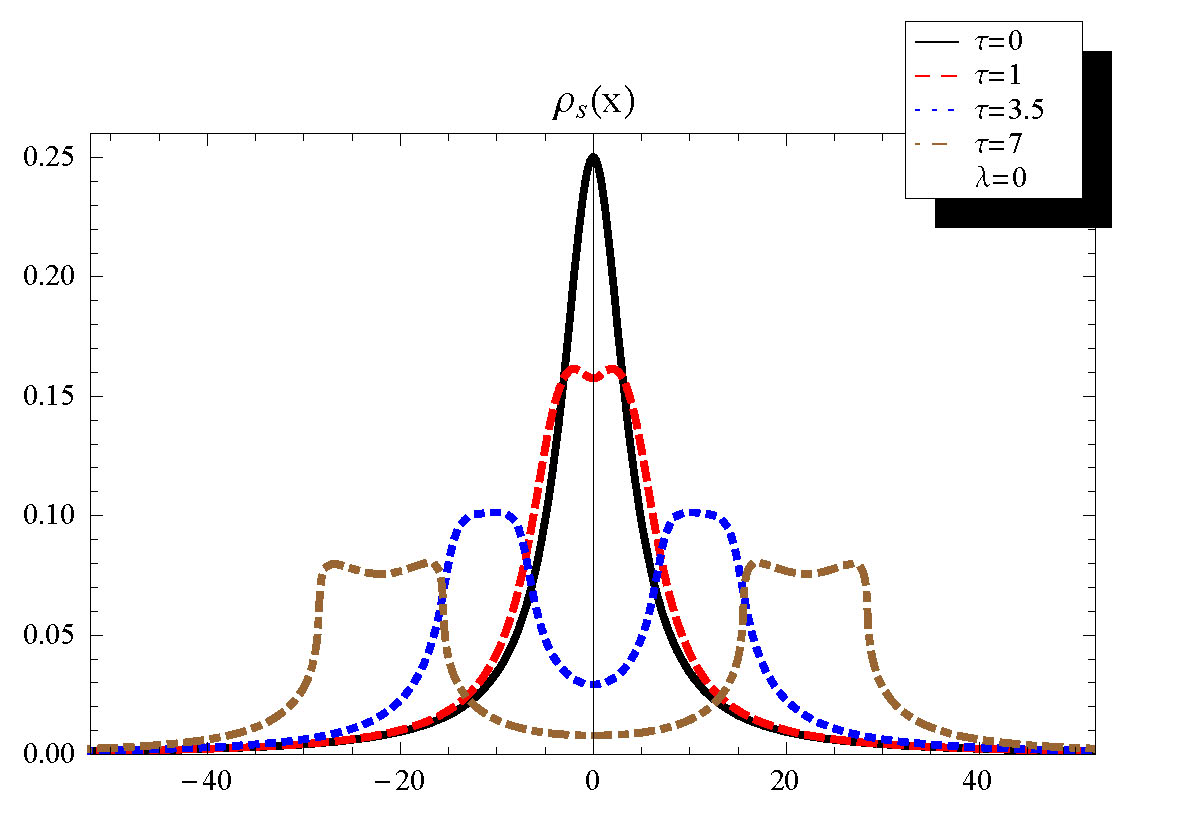}
\includegraphics[width=0.45\columnwidth]{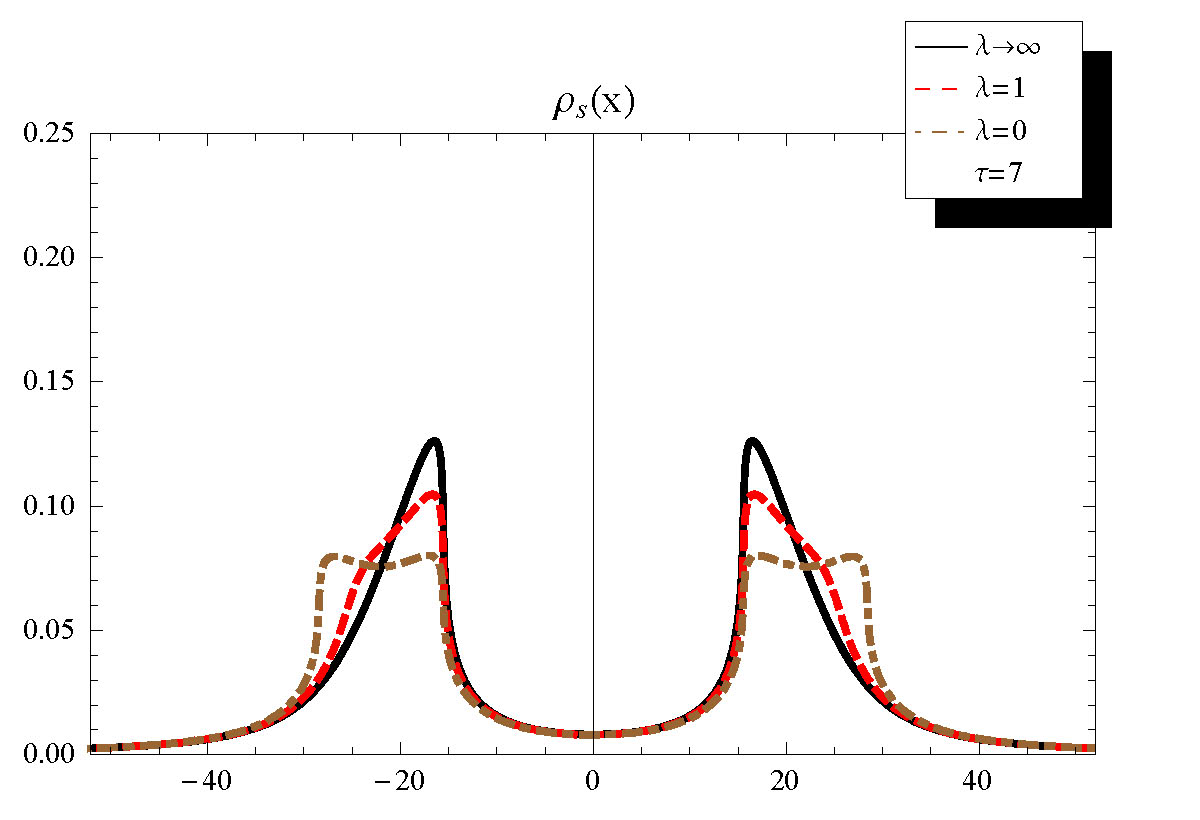}
\caption{
\label{fig:rhos_LO_lor}
 \textit{Left panel}: Spin dynamics of polarized center for free
fermions. The initial charge density profile is constant and the
initial spin density profile is a Lorentzian (\ref{initLor}) of a height $h=0.25$ and a half-width
$a=4$. Profiles at times $\tau=t/2=0, 1, 3.5, 7$ are shown. \textit{Right panel}: A snapshot of spin density at time $t=\tau/(\lambda+1/2)$ for $\tau=7$ for $\lambda=0,1,\infty$.}
\end{figure}

\begin{figure}
\includegraphics[width=0.45\columnwidth]{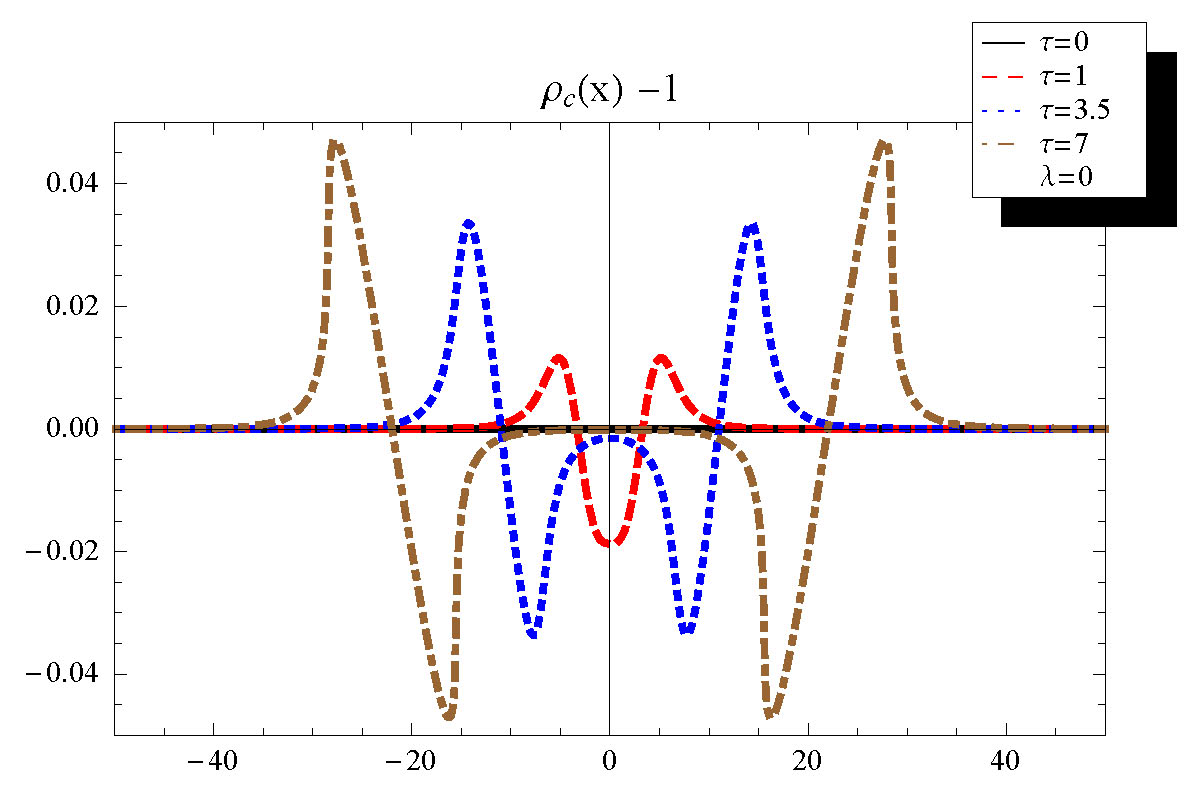}
\includegraphics[width=0.45\columnwidth]{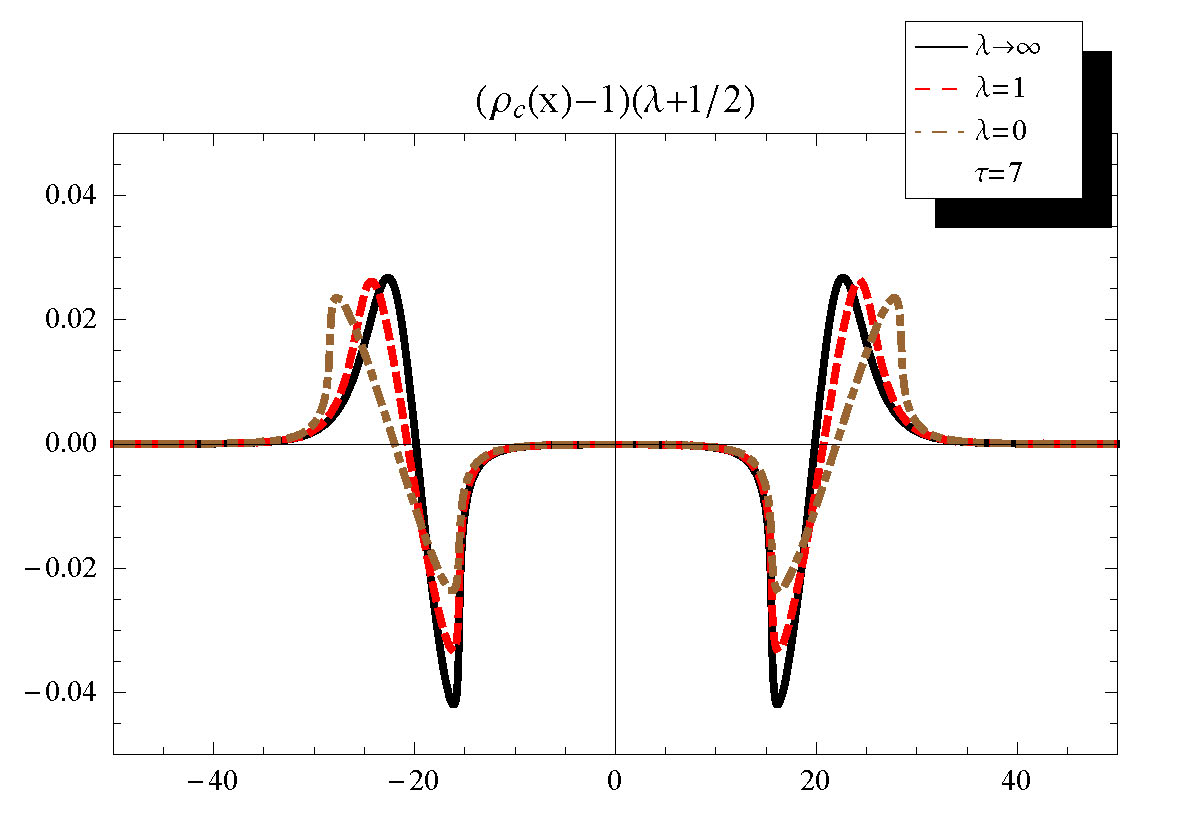}
\caption{
\label{fig:rhoc_LO_lor}
\textit{Left panel}:
Charge dynamics of polarized center for free
fermions. The initial charge density profile is constant and the
initial spin density profile is a Lorentzian (\ref{initLor}) of a height $h=0.25$ and a half-width
$a=4$. Profiles at times $\tau=t/2=0, 1, 3.5, 7$ are shown. \textit{Right panel}: A snapshot of a rescaled charge density $(\lambda+1/2)(\rho_{c}-1)$ at time $t=\tau/(\lambda+1/2)$ for $\tau=7$ for $\lambda=0,1,\infty$.}
\end{figure}

We present the results for spin and charge dynamics of free fermions with polarized center initial conditions (\ref{initLor}) on left panels of Figures \ref{fig:rhos_LO_lor},\ref{fig:rhoc_LO_lor}. The profiles $\rho_{s}(x)$ and $\rho_{c}(x)-1$ are shown as functions of $x$ for times $\tau=0, 1, 3.5, 7$ respectively. Here we use a rescaled time $\tau=(\lambda+1/2)t=t/2$ for future convenience.

The dynamics is separated into four Riemann-Hopf equations for each Fermi momenta. The initial conditions (\ref{initLor}) can be written as Lorentzian peaks for each of the four Fermi momenta of fermions and all four Fermi velocities are different. This results in a splitting of an initial Lorentzian peak into four peaks at larger times which can be easily seen on the left panel of Fig.~\ref{fig:rhos_LO_lor}. In addition to this linear effects the nonlinear effects of steepening the wave front can also be seen. The latter will render gradientless hydrodynamics inapplicable at later times.

The drag of charge by spin clearly seen in Fig.~\ref{fig:rhoc_LO_lor} has an essentially nonlinear nature. There is an excess (deficit) of particles with spin up (down) at the origin at the initial moment. The particles with spin up will move away from the center while spin down particles will move towards the center. However, the  average velocity of spin up particles is larger than the average velocity of spin down particles as it is proportional to the density of those particles. Therefore, the initial motion of particles away and towards the origin creates a charge depletion in the center and charge density maxima away from that depletion. This gives a qualitative explanation of the picture of charge dragged by spin which is shown in the left panel of Fig.~\ref{fig:rhoc_LO_lor}. Notice that in this explanation we used the dependence of propagation velocity on the amplitude of the wave -- an essentially nonlinear effect.

\subsubsection{\texorpdfstring{$\lambda$}{}-dependence of spin and charge dynamics}

To see the effects of the interaction on spin and charge dynamics we show the spin and charge density profiles at a fixed time for different values of the coupling constant $\lambda$ in the right panels of Figures \ref{fig:rhos_LO_lor},\ref{fig:rhoc_LO_lor} respectively. It is convenient to use the scaling dictated by the $\lambda\to \infty$ limit considered in detail in Section \ref{sec:Lattice}. Namely, we use a rescaled time $\tau=(\lambda+1/2)t$ and rescale the deviation of the charge
density from the uniform by background plotting $(\lambda+1/2)(\rho_{c}-1)$ for the charge density. The charge and density profiles found at $\tau=7$ are remarkably close for $\lambda$ ranging from the free fermion case $\lambda=0$ to the limit  of Haldane-Shastry model $\lambda\to \infty$.

The results confirm that the effect of spin dynamics on charge is suppressed by $1/\lambda$ for large $\lambda$. For a given initial spin density profile the maximal amplitude of charge deviation is of the order $1/(\lambda+1/2)$.

\section{Conclusions}
 \la{sec:concl}

In this paper we considered a classical two-fluid hydrodynamics derived as a semiclassical limit of the quantum spin-Calogero model (sCM) defined in (\ref{eq:h_afm}). The model (\ref{eq:h_afm}) is essentially quantum as it involves identical particles and a particle permutation operator. There is an essential ambiguity in how one takes a ``semiclassical'' limit. Here we considered a limit which is obtained when the density of particles goes to infinity so that $\hbar \rho$ is kept finite in the  limit $\hbar\to 0$. We have also neglected gradient corrections to hydrodynamic equations assuming that fields change very slowly on the scale of the inter-particle spacing. With all these assumptions, hydrodynamic equations are obtained from the Bethe ansatz solution of sCM. They have the simplest form when written in terms of fields corresponding to dressed Fermi momenta of Bethe ansatz. In terms of these fields (\ref{eq:uup},\ref{eq:udown}), equations separate into four independent Riemann-Hopf equations (\ref{RE}) which are trivially integrable.

We presented some particular solutions of the hydrodynamic equations illustrating interactions between spin and charge. There is no true spin-charge separation in sCM. However, in the limit of large coupling constant $\lambda\to\infty$ the spin degrees of freedom do not affect the dynamics of charge degrees of freedom. The spin dynamics then is described by the hydrodynamics of Haldane-Shastry spin model. We considered explicitly both this limit ($\lambda\to\infty$) and the limit ($\lambda=0$) of free fermions with spin.

The quantum scattering phase of particles interacting via $1/x^{2}$ potential is momentum independent. Moreover, it is the same for particles of the same species and for particles of different species because of the $SU(2)$ invariance of (\ref{eq:h_afm}). It is well known that this allows one to describe sCM as a model of free exclusons - particles obeying an exclusion statistics \cite{1991-Haldane-exclusion,1995-FukuiKawakami,1996-KatoKuramoto,1999-Polychronakos-LesHouches}. We do not keep the $SU(2)$ invariance of the original quantum model (\ref{eq:h_afm}) explicitly when taking the classical limit. However, this invariance is responsible for the variable separation that we observed in our hydrodynamics. We note here that sCM can be generalized to the ``multi-species Calogero model'' \cite{Bardek-two-family}. Because of the absence of the $SU(2)$ invariance for a more general two-species Calogero model one does not have the separation of variables for the corresponding hydrodynamics.

The classical gradientless hydrodynamics derived in this paper captures a lot of the features of sCM. It is straightforward to generalize our results to the case of the $SU(n)$ Calogero model and to use the gradientless hydrodynamic equations for problems where field gradients can be neglected. In a separate publication \cite{2009-Franchini-Kulkarni} we use these equations in instanton calculations for the computation of emptiness formation probability similar to what was done in Refs \cite{2005-Abanov-LesHouches, 2005-FranchiniAbanov}.

However, some important features of the hydrodynamic description do require an account of gradient corrections. First of all, the exact hydrodynamic equations are expected not to have an exact separation of variables. The obtained Riemann-Hopf equations (\ref{RE}) acquire gradient corrections and four such equations written for (\ref{eq:uup},\ref{eq:udown}) are expected to be coupled by those gradient corrections similarly to the case of the one-species Calogero-Sutherland model \cite{2009-AbanovBettelheimWiegmann}. Similarly, we expect that the equations with gradient corrections will have soliton solutions corresponding to quasi-particle excitations of the quantum model (\ref{eq:h_afm}) \cite{1995-Polychronakos,Sutherland-book,2009-AbanovBettelheimWiegmann}.

The hydrodynamic description of quantum sCM has been addressed in Refs.\cite{1996-AMY,jevicki-notes} using the collective field theory approach. The comparison of our results with results of those works is not straightforward. One should apply the collective formulation of Refs.\cite{1996-AMY,jevicki-notes} to the states from an appropriate sector of coherent states and take a corresponding classical limit. It would be especially interesting to see how the three hydrodynamic regimes discussed here appear from Refs.\cite{1996-AMY,jevicki-notes}. One can also recognize a lot of similar looking terms in quantum hydrodynamics of Refs. \cite{jevicki-notes} and in our classical gradientless hydrodynamics. It would also be very important to understand the role of the degeneracy due to the Yangian symmetry in sCM on its hydrodynamics. The latter degeneracy was neglected in the classical hydrodynamics of this paper.

\section{Acknowledgments}

We all have  benefited from discussions with A. Polychronakos.
A.G.A. is grateful to Y. Kato for a discussion of spin-charge (non)separation in spin-Calogero model. M.K. thanks D. Schneble for discussions of spin-charge dynamics in interacting systems. A.G.A. and F.F. would like to acknowledge the hospitality of the Galileo Galilei Center for Theoretical Physics (Florence, Italy) during the workshop on {\it Low-dimensional Quantum Field Theories and Applications}, where some interesting discussions took place.
The work of A.G.A. was supported by the NSF under the grant DMR-0348358.

\appendix

\section{Asymptotic Bethe Ansatz solution of spin Calogero Model and separation of variables in hydrodynamics}
 \label{app:AABsCM}

The spin Calogero model is solvable by asymptotic Bethe Ansatz (ABA)\cite{1992-Kawakami,sutherland-shastry-1993-PRL,Sutherland-book}. This solution turns out to be the most convenient for our purposes.

The most important ingredient of ABA is the scattering phase which is given by
\be
	\theta(k)=\pi \lambda\, \sign(k)
   \label{sCMtheta}
\ee
for sCM. Here $k$ is the relative momentum of two particles and the scattering phase does not depend on the species of particles. The expression for the dressed (true physical) momentum of the particle is given by
\be
	k(\kappa) = \frac{2\pi}{L}\left[\kappa
	 +\frac{\lambda}{2}\int_{-\infty}^{\infty}\sign(\kappa-\kappa')\nu(\kappa')\,\de \kappa'\right],
 \label{dmomentum}
\ee
where $\kappa$ is an integer-valued non-interacting momentum of the particle (quantum number) and $\nu(\kappa)$ is the number of particles with quantum number $\kappa$ (see (\ref{eq:nutotal})). Here we replaced in the scattering phase $\sgn(k-k')$ by $\sgn(\kappa-\kappa')$\cite{monotonicity}.

\begin{figure}
   \includegraphics[width=6cm]{co-phasespace.jpg}\includegraphics[width=6cm]{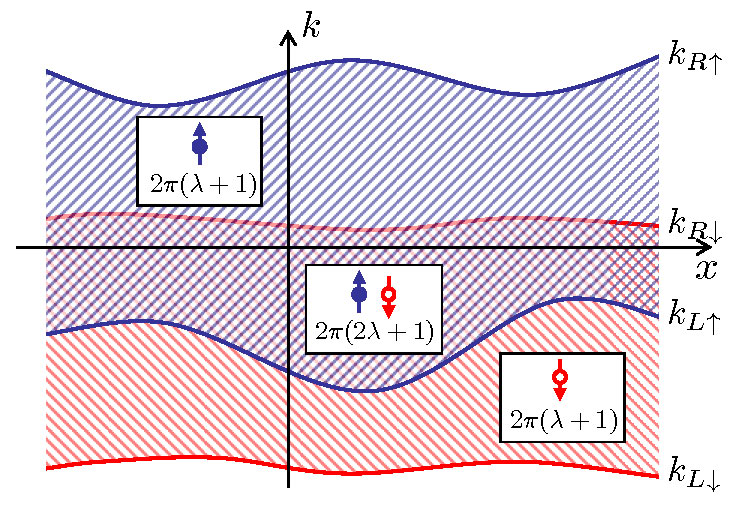}\includegraphics[width=6cm]{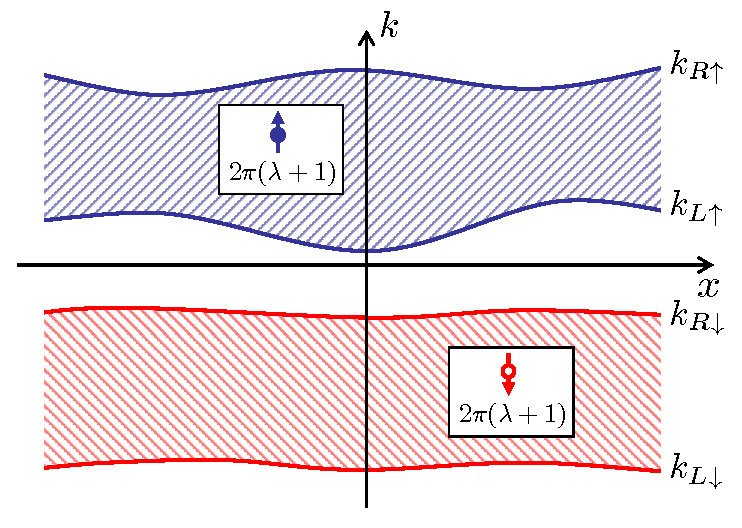}
   \caption{\label{fig:phase-space-all} Phase-space diagrams of a hydrodynamic states characterized by four space-dependent Fermi momenta in three regimes CO, PO, and NO respectively.}
\end{figure}

We immediately obtain from (\ref{dmomentum})
\be
	L \frac{\de k}{2\pi} = \left[1+\lambda \nu(\kappa)\right]\, \de \kappa
\ee
and
\be
	\nu_{\uparrow (\downarrow)}(\kappa)\, \de \kappa
	=L \frac{\nu_{\uparrow (\downarrow)}(k)}{1+\lambda \nu(k)}\, \frac{\de k}{2\pi}.
 \la{nudkappa}
\ee
We can see that the  picture corresponding to (\ref{nudkappa}) in a single-particle phase space requires that the number of particles in the phase space volume is given by $\frac{dx\,dk}{2\pi(\lambda+1)}$ if only one species is present and $\frac{dx\,dk}{2\pi(\lambda+1/2)}$ when both species are present. This justifies the picture we used (see Figs. \ref{fig:phase-space} and \ref{fig:phase-space-all}).

It is easy to write down the expressions for the conserved quantities using (\ref{nudkappa}).
\bea
	N_{\uparrow (\downarrow)} &=& L\int_{-\infty}^{+\infty}\frac{\de k}{2\pi}\,
	\frac{\nu_{\uparrow (\downarrow)}(k)}{1+\lambda \nu(k)}\;
 \la{Ndmomenta} \\
	P &=& L\int_{-\infty}^{+\infty}\frac{\de k}{2\pi}\,
	\frac{\nu(k)}{1+\lambda \nu(k)}\; k
 \la{Pdmomenta} \\
	P_{s} &=& L\int_{-\infty}^{+\infty}\frac{\de k}{2\pi}\,
	\frac{\nu_{s}(k)}{1+\lambda \nu(k)}\; k
 \la{Psdmomenta} \\
	E &=& L\int_{-\infty}^{+\infty}\frac{\de k}{2\pi}\,
	\frac{\nu(k)}{1+\lambda \nu(k)}\; \frac{k^{2}}2.
 \la{Edmomenta}
\eea
Here $P_{s}$ is a conserved quantity proportional to ${\cal L}_{1}^{z}$ introduced in \cite{1993-HikamiWadati-conserved}:
\bea
   \hat{P}_s \equiv
   - \ii \sum_{j=1}^N \sigma_j^z \frac{\partial}{\partial x_j}
   - \ii \frac{\lambda}{2} \; \frac{\pi}{L} \sum_{j \ne l} \cot \frac{\pi}{L} \left( x_j - x_l \right) \; \left[ \sigma_j^z - \sigma_l^z \right] {\rm P}_{jl} \; .
\eea

One can think of, e.g.,  $P_{\uparrow}=(P+P_{s})/2$ as of a sum of asymptotic values of momenta of spin-up particles.
We have replaced summations by integrations as we need only continuous versions of these formulae.
It can be shown that (\ref{Ndmomenta},\ref{Pdmomenta},\ref{Edmomenta}) are equivalent to (\ref{eq:N},\ref{eq:momentum},\ref{eq:eigenE}) with the relation between physical and non-interacting momenta given by (\ref{dmomentum}). Moreover, because the measure of integration $\frac{dk}{2\pi}\,
\frac{\nu_{\uparrow (\downarrow)}(k)}{1+\lambda \nu(k)}$ is a piece-wise constant for the two-step distribution (\ref{eq:nugs}), one naturally obtains integrals of motion in a form which is completely separated in terms of Fermi momenta. Indeed for a two-step distribution (\ref{eq:nugs})
\bea
	\nu_{\alpha} = \left\{\begin{array}{ll}
							1, & \mbox{  if  }k_{L\alpha}<k<k_{R\alpha} \\
							0, & \mbox{  otherwise}
						\end{array}\right.
\eea
where $\alpha=\uparrow,\downarrow$. In the CO regime (\ref{COdmomenta}) we have
\bea
	\int_{-\infty}^{\infty}\frac{\de k}{2\pi} \,
	\frac{\nu_{\uparrow}(k)}{1+\lambda \nu(k)}\; f(k)
	& = &
	\int_{k_{L\uparrow}}^{k_{L\downarrow}}\frac{\de k}{2\pi} \,
	\frac{1}{1+\lambda}\; f(k)	
	+\int_{k_{L\downarrow}}^{k_{R\downarrow}}\frac{\de k}{2\pi} \,
	\frac{1}{1+2\lambda}\; f(k)	
	+\int_{k_{R\downarrow}}^{k_{R\uparrow}}\frac{\de k}{2\pi} \,
	\frac{1}{1+\lambda}\; f(k),
 \nonumber \\
	\int_{-\infty}^{\infty}\frac{\de k}{2\pi} \,
	\frac{\nu_{\downarrow}(k)}{1+\lambda \nu(k)}\; f(k)
	& = &
	\int_{k_{L\downarrow}}^{k_{R\downarrow}}\frac{\de k}{2\pi} \,
	\frac{1}{1+2\lambda}\; f(k),
\eea
where $f(k)$ is an arbitrary function. In particular, we obtain for the densities
\bea
	2\pi (\lambda+1) \; \frac{N}{L}
	&=& k_{R\uparrow}-k_{L\uparrow}
	+\frac{1}{2\lambda+1}(k_{R\downarrow}-k_{L\downarrow}),
 \la{Nfermi} \\
	2\pi (\lambda+1) \; \frac{N_{s}}{L}
	&=& k_{R\uparrow}-k_{L\uparrow}
	-(k_{R\downarrow}-k_{L\downarrow}),
 \la{Nsfermi} \\
 	4\pi (\lambda+1) \; \frac{P}{L}
	&=& k_{R\uparrow}^{2}-k_{L\uparrow}^{2}
	+\frac{1}{2\lambda+1}(k_{R\downarrow}^{2}-k_{L\downarrow}^{2}),
 \la{Pfermi}
 \\
 	4\pi (\lambda+1) \; \frac{P_{s}}{L}
	&=& k_{R\uparrow}^{2}-k_{L\uparrow}^{2}
	-(k_{R\downarrow}^{2}-k_{L\downarrow}^{2}),
 \la{Psfermi}
 \\
 	12\pi (\lambda+1) \; \frac{E}{L}
	&=& k_{R\uparrow}^{3}-k_{L\uparrow}^{3}
	+\frac{1}{2\lambda+1}(k_{R\downarrow}^{3}-k_{L\downarrow}^{3}).
 \la{Efermi}
\eea

So far we presented the values of the conserved quantities for the sCM in terms of dressed Fermi momenta. They are given by linear combinations of Fermi momenta raised to the same power. There are infinitely many integrals of motion of this type and they are all in involution (commute with each other). The latter is a pretty stringent requirement and we assume that the only way to satisfy it is to require that the corresponding classical hydrodynamic fields have the following Poisson's brackets
\be
	 \left\{k_{\alpha}(x), k_{\beta}(y) \right\}= 2\pi s_{\alpha} \delta_{\alpha\beta} \delta'(x-y),
 \la{ksalpha}
\ee
where $\alpha$ runs over all Fermi points and $s_{\alpha}$ are some numbers to be determined. We can determine these numbers, e.g., in the following way. The density of current $j$ (momentum per unit length) from (\ref{Pfermi}) by
\be
	j(x) = \frac{1}{4\pi (\lambda+1)}\left[k_{R\uparrow}^{2}-k_{L\uparrow}^{2}
	+\frac{1}{2\lambda+1}(k_{R\downarrow}^{2}-k_{L\downarrow}^{2})\right].
 \la{cdensfermi}
\ee
The total momentum of the system is a generator of the translation algebra $\{P,q(y)\}=\partial_{y}q(y)$, where $q(y)$ is any field. For the current density we should have
\be
	\left\{j(x), q(y)\right\} = q(x)\delta'(x-y).
 \la{currentPB}
\ee
Taking $q(y)$ to be $k_{\alpha}(y)$ and combining (\ref{currentPB}) with (\ref{ksalpha}) we
fix the unknown coefficients $s_{\alpha}$
\bea
	s_{R\uparrow} &=& -s_{L\uparrow}=\lambda+1,
 \nonumber \\
 	s_{R\downarrow} &=& -s_{L\downarrow}= (\lambda+1)(2\lambda+1).
 \la{salpha}
\eea
Computing Poisson's bracket of the hydrodynamic Hamiltonian (obtained from (\ref{Efermi})
\be
	H = \frac{1}{12\pi (\lambda+1)} \int \de x\, \left[
	k_{R\uparrow}^{3}-k_{L\uparrow}^{3}
	+\frac{1}{2\lambda+1}(k_{R\downarrow}^{3}-k_{L\downarrow}^{3})\right]
 \la{Hdm}
\ee
with $k_{\alpha}(x)$ we obtain Riemann-Hopf equation (\ref{RE}) for every Fermi momentum field $k_{\alpha}(x,t)$.

\section{Hydrodynamic velocities}
 \la{app:velocities}

In appendix \ref{app:AABsCM} we did not use the notion of hydrodynamic velocity. Instead, our hydrodynamic equations were written directly in terms of dressed Fermi momentum fields $k_{\alpha}(x,t)$. We also know how to express other quantities like density, momentum, energy, etc in terms of these variables. Let us now find the expressions for the velocity fields $v_{\uparrow,\downarrow}$. We focus on the CO regime here and consider other regimes in appendix \ref{app:allcases}.

First of all we, give the expressions for the  conserved densities and conserved current densities  which can be found from (\ref{Nfermi},\ref{Nsfermi},\ref{Pfermi},\ref{Psfermi}) as
\bea
	\rho_{\uparrow} &=& \frac{\rho+\rho_{s}}{2}
	= \frac{1}{2\pi(\lambda+1)} \left[k_{R\uparrow}-k_{L\uparrow}
	-\frac{\lambda}{2\lambda+1}(k_{R\downarrow}-k_{L\downarrow})\right],
 \nonumber \\
	\rho_{\downarrow} &=& \frac{\rho-\rho_{s}}{2}
	= \frac{1}{2\pi(2\lambda+1)} (k_{R\downarrow}-k_{L\downarrow}),
 \la{rhoupdown} \\
	j_{\uparrow} &=& \frac{j+j_{s}}{2}
	= \frac{1}{4\pi(\lambda+1)} \left[k_{R\uparrow}^{2}-k_{L\uparrow}^{2}
	 -\frac{\lambda}{2\lambda+1}(k_{R\downarrow}^{2}-k_{L\downarrow}^{2})\right],
 \nonumber \\
	j_{\downarrow} &=& \frac{j-j_{s}}{2}
	= \frac{1}{2\pi(2\lambda+1)} (k_{R\downarrow}^{2}-k_{L\downarrow}^{2}).
 \la{jupdown}
\eea
In hydrodynamics, the velocities are defined as variables conjugated to the conserved momenta. Namely, the differential of the energy density defines chemical potentials and velocities as
\be
	\de \epsilon = \mu_{\uparrow} \de \rho_{\uparrow}
    + \mu_{\downarrow} \de \rho_{\downarrow}
	+v_{\uparrow}^{h} \de j_{\uparrow}
    +v_{\downarrow}^{h} \de j_{\downarrow}.
 \la{hveldef}
\ee
Using the energy density obtained from (\ref{Hdm}) we have
\be
   \de \epsilon = \frac{1}{4\pi (\lambda+1)} \left[
   k_{R\uparrow}^{2} \; \de k_{R\uparrow}
   - k_{L\uparrow}^{2} \; \de k_{L\uparrow}
   + \frac{1}{2\lambda+1} \left(
   k_{R\downarrow}^{2} \; \de k_{R\downarrow}
   - k_{L\downarrow}^{2} \; \de k_{L\downarrow} \right) \right]
\ee
and using (\ref{rhoupdown},\ref{jupdown}) one can determine $\mu_{\uparrow,\downarrow}$ and $v_{\uparrow, \downarrow}$. The hydrodynamic velocities are given by linear combinations of Fermi momenta \cite{linear-note}
\bea
	v_{\uparrow}^{h} &=& \frac{1}{2}(k_{R\uparrow}+k_{L\uparrow}),
 \nonumber \\
	v_{\downarrow}^{h} &=& \frac{1}{2(\lambda+1)}
	\left[\lambda(k_{R\uparrow}+k_{L\uparrow})
	+(k_{R\downarrow}+k_{L\downarrow})\right].
 \la{vupdown}
\eea
Using (\ref{ksalpha},\ref{salpha}) one can check that the velocities (\ref{vupdown}) have canonical Poisson's brackets with densities (\ref{rhoupdown})\cite{linear-note}
\be
	\{\rho_{\alpha}(x),v_{\beta}(y)\}=\delta_{\alpha\beta}\delta'(x-y),
\ee
where $\alpha,\beta=\uparrow,\downarrow$. The other Poisson's brackets vanish.

The hydrodynamic velocities (\ref{vupdown}) are precisely the ones used in the main body of this paper for CO regime $v_{\uparrow,\downarrow}=v_{\uparrow,\downarrow}^{h}$. Equations (\ref{eq:uup},\ref{eq:udown}) are the inverse to (\ref{rhoupdown},\ref{vupdown}). Interestingly, in the CO regime the velocities and densities of different species can be naturally (simply) written in terms of bare non-interacting momenta (\ref{eq:kup},\ref{eq:k3kdown}).

The current density in terms of densities and velocities follows from  (\ref{cdensfermi}) (compare with (\ref{eq:momentum_g}))
\be
	j(x)=\rho_{\uparrow}v_{\uparrow}+\rho_{\downarrow}v_{\downarrow}.
\ee
The density of ``spin-current'' which follows from (\ref{Psfermi}) has a ``correction'' proportional to $\lambda$ compared to the case of free fermions
\be
	j_{s}(x)=\rho_{\uparrow}v_{\uparrow}-\rho_{\downarrow}v_{\downarrow}
	+2\lambda \rho_{\downarrow}(v_{\uparrow}-v_{\downarrow}).
\ee

In this appendix we focused on CO regime. Of course, the formalism reviewed here is applicable to all three hydrodynamic regimes (CO, PO, and NO). We collect appropriate results in Appendix \ref{app:allcases}.

\section{Hydrodynamic regimes for spin-Calogero model}
 \la{app:allcases}

Depending on the relative order of four quantum numbers $\kappa_{R,L;\uparrow,\downarrow}$ we distinguish six different hydrodynamic regimes of the sCM. These regimes can be reduced to three essentially different ones exchanging  $\uparrow \; \leftrightarrow \; \downarrow$. In this appendix we consider these three regimes and then combine all six cases.

Before we proceed, let us remark that the function $k(\kappa)$ defined in (\ref{dmomentum}) is monotonic and  the order of the  quantum numbers $\kappa_{R,L;\uparrow,\downarrow}$ is the same as the one of the physical dressed momenta $k_{R,\uparrow}=k(\kappa_{R,\uparrow})$ etc. Therefore, we can use the latter to define hydrodynamic regimes instead of the bare momenta $\kappa$.

\subsection{Conserved densities and dressed Fermi momenta}

Let us consider generally some integrable system which has two infinite families of mutually commuting conserved quantities. We assume further that the densities of these quantities are given in terms of four dressed Fermi momenta $k_{\alpha}(x)$ with $\alpha=1,2,3,4$ as
\bea
	j_{n}(x) &=& \frac{1}{n}\sum_{\alpha=1}^{4}a_{\alpha}(k_{\alpha}(x))^{n},
 \nonumber \\
	j^{s}_{n}(x) &=& \frac{1}{n}\sum_{\alpha=1}^{4}a_{\alpha}b_{\alpha}(k_{\alpha}(x))^{n}.
 \la{consdens}
\eea
Here $n=1,2,3,\ldots$ and $a_{\alpha},b_{\alpha}$ are constant coefficients. We assumed that the conserved densities can be expressed locally in terms of $k_{\alpha}$ and neglected gradient corrections.

We identify the first several integrals with densities, currents, and the energy as
\bea
	j_{1}(x) &=& \rho(x),
 \nonumber \\
 	j_{1}^{s}(x) &=& \rho_{s}(x),
 \nonumber \\
 	j_{2}(x) &=& j(x),
 \nonumber \\
 	j_{2}^{s}(x) &=& j_{s}(x),
 \nonumber \\
 	j_{3}(x) &=& 2\epsilon(x).
 \la{ident}
\eea
We notice here that due to (\ref{Ndmomenta}-\ref{Edmomenta}) the identifications (\ref{ident}) (with (\ref{consdens})) are valid for sCM model in all its regimes. The higher order conserved densities (\ref{consdens}) correspond to conserved quantities of sCM introduced in Ref.\cite{1993-HikamiWadati-conserved}.

The requirement of vanishing Poisson's brackets between conserved quantities is very restrictive. It can be resolved by requiring canonical Poisson's brackets between Fermi momenta (\ref{ksalpha}). If (\ref{ksalpha}) is valid, it is easy to check that $\{\int dx\,j_{n}(x),\int dy\,j_{m}^{s}(y)\}=0$ etc. Using the fact that the total current is the generator of translations (\ref{currentPB}) we can fix the coefficients $s_{\alpha}$ in (\ref{ksalpha}) as  $2\pi s_{\alpha}=1/a_{\alpha}$ and obtain
\be
	 \{k_{\alpha}(x),k_{\beta}(y)\}=\frac{1}{a_{\alpha}}\delta_{\alpha\beta}\delta'(x-y).
 \la{kaalpha}
\ee
Using the Poisson's brackets (\ref{kaalpha}) and the Hamiltonian $H=\int dx\, \epsilon(x)$ with (\ref{ident},\ref{consdens}) it is easy to obtain the Riemann-Hopf evolution equations for the dressed Fermi momenta
\be
	\partial_{t}k_{\alpha}+k_{\alpha}\partial_{x}k_{\alpha} =0, \qquad \mbox{for  }\alpha=1,2,3,4
 \la{RHapp}
\ee
and the evolution equations for all conserved densities as
\bea
	\partial_{t}j_{n}+\partial_{x}j_{n+1} &=& 0,
 \nonumber \\
 	\partial_{t}j_{n}^{s}+\partial_{x}j_{n+1}^{s} &=& 0.
 \la{eqmotjn}
\eea
In the hydrodynamic regime only four of the densities are algebraically independent (as there are only four dressed Fermi momenta). Therefore, one can find constitutive relations, i.e., express the energy density in terms of $\rho, \rho_{s}, j$ and $j_{s}$. Alternatively, one can use hydrodynamic velocities $v^{h}$ and $v_{s}^{h}$ defined by (\ref{hveldef}) instead of currents $j, j_{s}$.

We can see that the hydrodynamics (\ref{consdens},\ref{ident},\ref{kaalpha}) is fully defined by coefficients $a_{\alpha},b_{\alpha}$. In fact, these coefficients are not totally independent. Requiring that densities $\rho$ and $\rho_{s}$ have vanishing Poisson's brackets with themselves and with each other gives three relations between the coefficients
\bea
	\sum_{\alpha}a_{\alpha} &=& 0,
 \nonumber \\
 	\sum_{\alpha}b_{\alpha}a_{\alpha} &=& 0,
 \nonumber \\
 	\sum_{\alpha}b_{\alpha}^{2}a_{\alpha} &=& 0.
 \la{abrel}
\eea
For CO, PO, and NO regimes of sCM these coefficients are summarized in the Table \ref{table:ab}. These coefficients do satisfy relations (\ref{abrel}).

\begin{table}
\centering
	\caption{Summary of three regimes. }\label{table:ab}
\begin{tabular}{|c|c|c|c|c|c|c|} \hline
 & $\alpha$ & $L\uparrow$ & $R\uparrow$ & $L\downarrow$ & $R\downarrow$
 &  \\ \hline
\multirow{2}{*}{\;CO\;}
& \;$2\pi(\lambda+1)\,a_{\alpha}$\;  & $\qquad-1\qquad $& $\qquad \;1\;\qquad$ & $-\frac{1}{2\lambda+1}$ & $\frac{1}{2\lambda+1}$ & \multirow{2}{*}{\;$k_{L\uparrow} < k_{L\downarrow} < k_{R\downarrow} < k_{R\uparrow}$\;} \\
& $b_{\alpha}$ & $1$ & $1$ & $-(2\lambda+1)$ & $2\lambda+1$ & \\ \hline
\multirow{2}{*}{PO}
& $2\pi(\lambda+1)\,a_{\alpha}$ & $-1$ & $\frac{1}{2\lambda+1}$ & $-\frac{1}{2\lambda+1}$ & $1$ & \multirow{2}{*}{\;$k_{L\downarrow} < k_{L\uparrow} < k_{R\downarrow} < k_{R\uparrow}$\;} \\
& $b_{\alpha}$ & $1$ & $2\lambda+1$ & $-(2\lambda+1)$ & $-1$ & \\  \hline
\multirow{2}{*}{NO}
& $2\pi(\lambda+1)\,a_{\alpha}$ & $-1$ & $1$ & $-1$ & $1$ & \multirow{2}{*}{\;$k_{L\downarrow} < k_{R\downarrow} < k_{L\uparrow} < k_{R\uparrow}$\;}\\
& $b_{\alpha}$ & $1$ & $1$ & $\qquad -1\qquad$ & $\qquad-1\qquad$ & \\  \hline
\end{tabular}
\end{table}

The matrix of Poisson's brackets of the dressed Fermi momenta $k_{\alpha}$ (\ref{kaalpha}) is diagonal but not proportional to the unit matrix. It is interesting that the Poisson's brackets of bare momenta $\kappa_{\alpha}$ satisfy
\be
	\{\kappa_{\alpha}(x),\kappa_{\beta}(y)\} = (-1)^{\alpha}\frac{L^{2}}{2\pi}\delta_{\alpha\beta}\delta'(x-y).
\ee
One then obtains that the ``velocities'' introduced in (\ref{eq:kup},\ref{eq:k3kdown}) are canonically conjugate to the corresponding densities and can be written as linear combinations of $\kappa_{\alpha}$ (and of $k_{\alpha}$). The velocities (\ref{eq:kup},\ref{eq:k3kdown}) are  defined just as conjugate variables to the densities. This definition is not unique. One can always shift $v_{\uparrow} \to v_{\uparrow}+2\pi \gamma \rho_{\downarrow}$ and $v_{\downarrow} \to v_{\downarrow}-2\pi \gamma \rho_{\uparrow}$ with any number $\gamma$ without changing Poisson's brackets.
The particular choice of variables (\ref{eq:kup},\ref{eq:k3kdown}) is convenient because it defines velocities continuously across all hydrodynamic regimes. Moreover, we have
\bea
	v_{\uparrow,\downarrow} &=& v^{h}_{\uparrow,\downarrow}, \hspace{2.1cm} \mbox{for CO},
 \nonumber \\
 	v_{\uparrow,\downarrow} &=& v^{h}_{\uparrow,\downarrow}
	\pm \pi \lambda \rho_{\downarrow,\uparrow}, \qquad \mbox{for NO}.
\eea
In PO regime  the hydrodynamic velocities are not linear combinations of $k_{\alpha}$ and their relations to the conjugated variables $v_{\uparrow,\downarrow}$ used in this paper are more complicated.

\subsection{Complete Overlap Regime (CO)}

The Complete Overlap regime corresponds to the case when
\be
   -\frac{\pi}{2} \; |\rho_s| < v_s < \frac{\pi}{2} \; |\rho_s| \; .
\ee
In this case the support of $\nu_{\downarrow}$ is a subset of the support of $\nu_{\uparrow}$ (or vice versa).
In the main body of the paper we mostly concentrated on this case, but for convenience we recap the main formulae in this appendix as well.
The dressed momenta (\ref{dmomentum}) in the CO regime for $\rho_s >0$, i.e., for the ordering
\be
   k_{L\uparrow} < k_{L\downarrow} < k_{R\downarrow} < k_{R\uparrow} \; ,
\ee
are
\begin{eqnarray}
	k_{R\uparrow,L\uparrow}
	& = & v_{\uparrow} \pm \pi\left[(\lambda+1)\rho_{\uparrow}+\lambda\rho_{\downarrow}\right]
	=v_{\uparrow}\pm \pi\rho_{\uparrow}\pm \lambda \pi \rho_{c},
 \nonumber\\
	k_{R\downarrow,L\downarrow}
	& = & (\lambda+1)v_{\downarrow}-\lambda v_{\uparrow}
	\pm \pi(2\lambda+1)\rho_{\downarrow}
	=v_{\downarrow}\pm \pi \rho_{\downarrow}+\lambda(-2v_{s}\pm 2\pi \rho_{\downarrow}).
 \label{eq:udownapp}
\end{eqnarray}

Poisson's brackets of $k_{\alpha}$ are given by (\ref{kaalpha}) with coefficients from the Table \ref{table:ab}. One can express all conserved densities (\ref{consdens}) in terms of dressed Fermi momenta using the Table \ref{table:ab}. For example, the Hamiltonian (see (\ref{ident})) reads
\bea
	H_{\rm CO} &=& \frac{1}{12\pi (\lambda+1)} \int \de x \, \left[
	k_{R\uparrow}^{3}-k_{L\uparrow}^{3}
	+\frac{1}{2\lambda+1}
    \left( k_{R\downarrow}^{3}-k_{L\downarrow}^{3} \right)\right]
 \la{HCOapp} \\
 	&=& \int \de x \; \Biggl\{
   \frac{1}{2} \rho_{\uparrow} v_{\uparrow}^{2}
   + \frac{1}{2} \rho_{\downarrow} v_{\downarrow}^{2}
   + \frac{\lambda}{2} \rho_{\downarrow}
   \big( v_{\uparrow} - v_{\downarrow} \big)^{2}
   \nonumber \\
   && \qquad \quad + \frac{\pi^{2}\lambda^{2}}{6} \rho_{c}^{3} +
   \frac{\pi^{2}}{6} \big(\rho_{\uparrow}^{3} + \rho_{\downarrow}^{3} \big) + \frac{\lambda\pi^{2}}{6} \big(2\rho_{\uparrow}^{3}
   + 3\rho_{\uparrow}^{2} \rho_{\downarrow}
   + 3\rho_{\downarrow}^{3} \big) \Biggr\}.
\eea

The evolution equations are given by (\ref{RHapp}) and can also be recast in terms of equations for densities and velocities (\ref{updownhydro},\ref{cshydro}).

\subsection{Partial Overlap Regime (PO)}

There are two regimes when the supports of $\nu_{\uparrow}$ and $\nu_{\downarrow}$ only partially overlap. Here we concentrate on the case for which
\be
   \frac{\pi}{2} \; |\rho_s| < v_s < \frac{\pi}{2} \; \rho_c \; ,
\ee corresponding to the ordering
\be
   k_{L\downarrow} < k_{L\uparrow} < k_{R\downarrow} < k_{R\uparrow} \; .
\ee
The other PO regime can be obtained by exchanging up and down particles, i.e. by changing $v_s \to - v_s$.
In this case the dressed momenta (\ref{dmomentum}) are
\begin{eqnarray}
	k_{L\downarrow} & = & v_{\downarrow}-\pi(\lambda+1)\rho_{\downarrow}
	-\pi\lambda\rho_{\uparrow}
	=v_{\downarrow}-\pi\rho_{\downarrow}-\lambda \pi\rho_{c},
 \nonumber\\
	k_{L\uparrow} & = & v_{\uparrow}+\lambda(v_{\uparrow}-v_{\downarrow})
	-\pi(2\lambda+1)\rho_{\uparrow}
	=v_{\uparrow}-\pi\rho_{\uparrow}+\lambda(2v_{s}-2\pi\rho_{\uparrow}),
 \nonumber\\
	k_{R\downarrow} & = & v_{\downarrow}-\lambda(v_{\uparrow}-v_{\downarrow})
	+\pi(2\lambda+1)\rho_{\downarrow}
	 =v_{\downarrow}+\pi\rho_{\downarrow}-\lambda(2v_{s}-2\pi\rho_{\downarrow}),
 \nonumber\\
	k_{R\uparrow} & = & v_{\uparrow}+\pi(\lambda+1)\rho_{\uparrow}
	+\pi\lambda\rho_{\downarrow}
	=v_{\uparrow}+\pi\rho_{\uparrow}+\lambda\pi\rho_{c}
\end{eqnarray}
and the Hamiltonian becomes (see Table \ref{table:ab} and (\ref{consdens},\ref{ident}))
\bea
	H_{\rm PO} &=& \frac{1}{12\pi (\lambda+1)} \int \de x \, \left[
	k_{R\downarrow}^{3}-k_{L\uparrow}^{3}
	+\frac{1}{2\lambda+1}
    \left(k_{R\uparrow}^{3}-k_{L\downarrow}^{3} \right)\right]
 \\
 	& = & \int \de x \Biggl\{
   \frac{1}{2}\rho_{\uparrow} v_{\uparrow}^{2}
   +\frac{1}{2}\rho_{\downarrow} v_{\downarrow}^{2}
   +\lambda\pi\rho_{\uparrow}\rho_{\downarrow}
   \left(v_{\downarrow}-v_{\uparrow}\right)
   -\frac{\lambda}{12\pi}
   \left[ v_{\uparrow}-v_{\downarrow} -\pi\left(\rho_{\uparrow}+\rho_{\downarrow}\right) \right]^{3}
   \nonumber \\
   && \qquad \quad + \frac{\pi^{2}\lambda^{2}}{6}\left(\rho_{\uparrow}+\rho_{\downarrow}\right)^{3}
   +\frac{\pi^{2}}{6}\left(1+2\lambda\right)\left(\rho_{\uparrow}^{3}+\rho_{\downarrow}^{3}\right)\Biggr\}.
\eea
Poisson's brackets of $k_{\alpha}$ are given by (\ref{kaalpha}) with coefficients from the Table \ref{table:ab} and evolution equations are given by (\ref{RHapp}).

\subsection{No Overlap Regime (NO)}

In this case, the supports of $\nu_{\uparrow}$ and $\nu_{\downarrow}$ do not overlap at all. For $v_s >0$ the ordering of dressed Fermi momenta is
\be
   k_{L\downarrow} < k_{R\downarrow} < k_{L\uparrow} < k_{R\uparrow}
\ee
and momenta themselves are
\begin{eqnarray}
   k_{R\uparrow,L\uparrow} & = & v_{\uparrow}+\pi\lambda\rho_{\downarrow}
   \pm \pi(\lambda+1)\rho_{\uparrow}
   =v_{\uparrow}\pm \pi \rho_{\uparrow}\pm \lambda \pi \rho_{c,s},
 \nonumber \\
  k_{R\downarrow,L\downarrow} & = & v_{\downarrow}-\pi\lambda\rho_{\uparrow}
  \pm \pi(\lambda+1)\rho_{\downarrow}
  =v_{\downarrow}\pm \pi\rho_{\downarrow}-\lambda \pi\rho_{s,c}.
\end{eqnarray}

and the Hamiltonian becomes (see Table \ref{table:ab} and (\ref{consdens},\ref{ident}))
\bea
	H_{\rm NO} &=& \frac{1}{12\pi (\lambda+1)} \int \de x\, \left[
	k_{R\uparrow}^{3}-k_{L\uparrow}^{3}
	+k_{R\downarrow}^{3}-k_{L\downarrow}^{3} \right]
 \\
 	& = & \int \de x \Biggl\{
   \frac{1}{2} \rho_{\uparrow}v_{\uparrow}^{2}
   +\frac{1}{2}\rho_{\downarrow}v_{\downarrow}^{2}
   +\lambda\pi\rho_{\uparrow}\rho_{\downarrow}\left(v_{\uparrow}-v_{\downarrow}\right)
   \nonumber \\
   && \qquad \quad +    \frac{\pi^{2}}{6}\left(\lambda+1\right)^{2}\left(\rho_{\uparrow}+\rho_{\downarrow}\right)^{3}
   -\frac{\pi^{2}}{2}\left(1+2\lambda\right)\rho_{\uparrow}\rho_{\downarrow}\left(\rho_{\uparrow}+\rho_{\downarrow}\right)
   \Biggl\}.
\eea
Poisson's brackets of $k_{\alpha}$ are given by (\ref{kaalpha}) with coefficients from the Table \ref{table:ab} and evolution equations are given by (\ref{RHapp}).

\subsection{All cases combined}

It is possible to combine all hydrodynamic regimes into relatively compact expressions introducing absolute values of hydrodynamic fields. A general Hamiltonian valid for all regimes takes a form
\begin{eqnarray}
   H & = & \int \de x \Biggl\{\frac{1}{2} \rho_{\uparrow}v_{\uparrow}^{2}
   +\frac{1}{2}\rho_{\downarrow}v_{\downarrow}^{2}
   +\frac{\pi^{2}}{6}\left(\rho_{\uparrow}^{3} +\rho_{\downarrow}^{3}\right)
   +\frac{\pi^{2}}{6}\lambda^{2}\rho_{c}^{3}
   +\frac{\pi^{2}}{3}2\lambda(\rho_{\uparrow}^{3}+\rho_{\downarrow}^{3})
   \nonumber \\
   && \qquad \quad
   + \lambda \; \rho_{c}\xi_{1}\xi_{2}
   - \frac{\lambda}{3\pi}\left(|\xi_{1}|^{3}+|\xi_{2}|^{3}\right)
   +\frac{\lambda}{3\pi}\left[ |\chi_{1}|^{3} - \chi_{1}^{3}+|\chi_{2}|^{3}
    + \chi_{2}^{3} \right]\Biggr\},
   \label{eq:allcases}
\end{eqnarray}
where we introduced the following notations
\begin{eqnarray}
    \xi_{1,2} & \equiv & v_{s} \pm \frac{\pi}{2} \; \rho_{s} \; ,
    \label{eq:xi} \\
    \chi_{1,2} & \equiv & v_{s} \pm \frac{\pi}{2} \; \rho_{c} \; .
    \label{eq:chi}
\end{eqnarray}
The Hamiltonian (\ref{eq:allcases}) can be obtained from (\ref{eq:eigenE},\ref{eq:eps}) for the general case of a two-step distribution function $\nu_{\uparrow,\downarrow}(\kappa)$.
We collect in the Table \ref{tab:Classification-of-different-cases} the information necessary to go quickly from the general expression (\ref{eq:allcases}) to the particular ones valid in separate regimes (CO, PO or  NO).
\begin{table}
\centering{}\begin{tabular}{|c|c|c|c|c|c|c|c|c|}
\hline
$k$ inequality & $\; v_s \; $ & $\; \rho_s \; $ & $\xi_{1}=v_{s}+\frac{\pi}{2}\rho_{s}$  & $\xi_{2}=v_{s}-\frac{\pi}{2}\rho_{s}$  & $\chi_{1}=v_{s}+\frac{\pi}{2}\rho_{c}$  & $\chi_{2}=v_{s}-\frac{\pi}{2} \rho_{c}$ & $v_s$ inequality & Regime \tabularnewline
\hline
\hline
$k_{L\uparrow}<k_{R\uparrow}<k_{L\downarrow}<k_{R\downarrow}$  &
$-$  &  &  $-$  & $-$  & $-$  & $-$ &
$v_s < - \frac{\pi}{2}\rho_c$ & NO
\tabularnewline
\hline
$k_{L\uparrow}<k_{L\downarrow}<k_{R\uparrow}<k_{R\downarrow}$  &
$-$  &  &  $-$  & $-$  & $+$  & $-$ &
$-\frac{\pi}{2} \rho_c < v_s < - \frac{\pi}{2} |\rho_s|$ & PO
\tabularnewline
\hline
$k_{L\downarrow}<k_{L\uparrow}<k_{R\uparrow}<k_{R\downarrow}$  &
  & $-$ &  $-$  & $+$  & $+$  & $-$ &
$\frac{\pi}{2} \rho_s < v_s < - \frac{\pi}{2} \rho_s$ & CO
\tabularnewline
\hline
$k_{L\uparrow}<k_{L\downarrow}<k_{R\downarrow}<k_{R\uparrow}$  &
  & $+$ &  $+$  & $-$  & $+$  & $-$ &
$-\frac{\pi}{2} \rho_s < v_s < \frac{\pi}{2} \rho_s$ & CO
\tabularnewline
\hline
$k_{L\downarrow}<k_{L\uparrow}<k_{R\downarrow}<k_{R\uparrow}$  &
$+$  &  &  $+$  & $+$  & $+$  & $-$ &
$\frac{\pi}{2} |\rho_s| < v_s < \frac{\pi}{2} \rho_c$ & PO
\tabularnewline
\hline
$k_{L\downarrow}<k_{R\downarrow}<k_{L\uparrow}<k_{R\uparrow}$  &
$+$  &  &  $+$  & $+$  & $+$  & $+$ &
$\frac{\pi}{2} \rho_c < v_s$ & NO \tabularnewline
\hline
\end{tabular}\caption{\label{tab:Classification-of-different-cases}Classification of different regimes: $+$ indicates that the field takes positive values, $-$ that it is negative. A blank means that its sign is arbitrary.}
\end{table}

We can combine the evolution equations following from (\ref{eq:allcases}) in the spin/charge basis (\ref{spinchargebasis}) as
\begin{eqnarray}
	\dot{\rho}_c & = & -\partial_{x}\Bigl\{\rho_c v_c + \rho_s v_s \Bigr\},
 \label{eq:rhocdot} \\
	\dot{\rho}_s & = & -\partial_{x}\Biggl\{\rho_c v_s + \rho_s v_c
   	- \frac{\lambda}{\pi}\Bigl[ \xi_{1} |\xi_{1}| + \xi_{2} |\xi_{2}|
   	- \chi_{1} |\chi_{1}| - \chi_{2} |\chi_{2}| \Bigr]\Biggr\},
 \label{eq:rhosdot} \\
   	\dot{v}_c & = & -\partial_{x}\Biggl\{\frac{v_c^2 + v_s^2}{2}
   	+\frac{\pi^{2}}{8}\Bigl[ \left( 4 \lambda^{2} + 2 \lambda + 1 \right) \rho_{c}^{2}
	+ \left( 2 \lambda + 1 \right) \rho_s ^{2}\Bigr]
	+ \frac{\lambda}{2}\Bigl[ \chi_{1} |\chi_{1}| - \chi_{2} |\chi_{2}| \Bigr]\Biggr\},
 \label{eq:vcdot} \\
   	\dot{v}_s & = & -\partial_{x}\Biggl\{ v_c v_s
   	+\frac{\pi^{2}}{4} \left(2\lambda+1\right) \rho_c \rho_s
	-\frac{\lambda}{2}\Bigl[ \xi_{1} |\xi_{1}| - \xi_{2} |\xi_{2}| \Bigr]\Biggr\}.
 \label{eq:vsdot}
\end{eqnarray}

For CO and PO regimes the Hamiltonian (\ref{eq:allcases}) takes an especially simple form in terms of dressed momenta
\begin{eqnarray}
    H_{\rm CO \, \& \, PO} & = &
    \frac{1}{12\pi\left(2\lambda+1\right)} \int \de x
    \Biggr\{ k_{R\uparrow}^3 - k_{L\uparrow}^3 + k_{R\downarrow}^{3} - k_{L\downarrow}^{3}
    + \frac{\lambda}{\left(\lambda+1\right)} \Bigl[
    \left| k_{L\uparrow}^{3}-k_{L\downarrow}^{3} \right|
    +\left| k_{R\uparrow}^{3}-k_{R\downarrow}^{3} \right| \Bigr]\Biggr\} \; ,
    \label{eq:hhii}
\end{eqnarray}
which are related to density and velocity fields as
\begin{eqnarray}
  k_{R\uparrow,L\uparrow} & = & v_{\uparrow} \pm \pi(\lambda+1)\rho_{\uparrow}+\lambda\chi_{1,2}
  \mp \lambda |\xi_{1,2}|
  \; , \nonumber \\
  k_{R\downarrow,L\downarrow} & = & v_{\downarrow} \pm \pi(1 + \lambda)\rho_{\downarrow}-\lambda\chi_{2,1}
  \mp \lambda |\xi_{1,2}|
  \; . \label{eq:ULRdown}
\end{eqnarray}

As in the separate cases considered before, these momenta have canonical Poisson's brackets (\ref{ksalpha}) with
\bea
	s_{R\uparrow,R\downarrow} &=& (\lambda+1)\left[\lambda+1 \pm \lambda\sgn(\xi_{1})\right]
 \nonumber \\
	s_{L\uparrow,L\downarrow} &=& -(\lambda+1)\left[\lambda+1 \mp \lambda\sgn(\xi_{2})\right]
   \label{salphaCOPO}
\eea

and evolve independently according to the Riemann-Hopf equations (\ref{RHapp}).

\section{Hydrodynamic description of Haldane-Shastry model from its Bethe Ansatz solution}
 \la{app:hdHSM}

The Haldane-Shastry model (HSM) is a Heisenberg spin chain with long-ranged interaction defined by the Hamiltonian:
\begin{equation}
   H_{\rm HSM} =\frac{1}{2} \sum_{j < l}
   \frac{{\bf K}_{jl}}{d \left(j - l \right)^{2}} \; ,
   \label{eq:Haldane_shastry_app}
\end{equation}
where we $K_{jl}$ is the spin-exchange operator\footnote{Note that for fermions ${\bf P}_{jl} \; {\bf K}_{jl} = - 1$.}:
\be
   {\bf K}_{jl} = \frac{\vec{\sigma}_{j} \cdot \vec{\sigma}_{l} + 1}{2} \; ,
\ee
and $d(j) \equiv (N/\pi) \left| \sin \left( \pi j / N \right) \right|$ is the chord distance between two points on a lattice with $N$ sites and periodic boundary conditions.
The model (\ref{eq:Haldane_shastry_app}) has been introduced independently at the same time by Haldane\cite{1988-Haldane-HS} and by Shastry\cite{1988-Shastry-HS} and has been shown to be integrable.
The energy spectrum of the HSM is equivalent to that of the Calogero-Sutherland model at $\lambda =2$, but with a high degeneracy due to the Yangian symmetry \cite{1992-HaldaneEtAl,HaHaldane93}.

In this appendix we used the Bethe Ansatz solution \cite{1988-Haldane-HS, Haldane91, Habook} to construct a gradientless hydrodynamic description for the HSM similarly to what we have done for the sCM model in section \ref{sec:hydr} and appendix \ref{app:AABsCM}. To this end, we consider a state with $M$ overturned spins over an initial ferromagnetic configuration (say from up to down and $M<N/2$) and introduce $M$ integer quantum numbers $\kappa$'s to characterize the state in the Bethe Ansatz formalism. As before such state can be described by a distribution function $\nu(\kappa)=0,1$, depending on whether that quantum number is present or not in the BA solution. Following \cite{1988-Haldane-HS} we impose a condition on the integer numbers: $|\kappa| < (N - M -1)/2$. \footnote{This corresponds to having a single compact support of $\nu$ within a single Brillouin zone. Other regimes will require an analysis of umklapp processes\cite{1992-Kawakami} and will not be considered here.}

The scattering phase for the HSM is
\be
   \theta (k) = \pi \; \sign (k) \; ,
\ee
which corresponds to setting $\lambda =1$ into (\ref{sCMtheta})\footnote{This scattering phase is identical to the one in $\lambda =2$  bosonic Calogero-Sutherland model \cite{1988-Haldane-HS}.}. Please note that since we are considering a lattice model, the momentum is defined within the Brillouin zone: $- \pi  < k < \pi $, where we took the lattice spacing as unity.

At this point, all the derivations of appendix \ref{app:AABsCM} can be repeated step by step for the HSM just by setting everywhere $\lambda =1$, and remembering that the momentum is always defined modulo $2 \pi$.
In particular, the dressed momentum is
\be
   k (\kappa) = \frac{2 \pi}{L} \left[ \kappa + \frac{1}{2} \int \sgn (\kappa - \kappa') \tilde{\nu} (\kappa') \de \kappa' \right],
\ee
where again we replaced $\sgn(k-k')$ by $\sgn(\kappa-\kappa')$\cite{monotonicity}
and the distribution of the physical momenta is given by
\be
   \nu (\kappa) \de \kappa =
   \frac{N}{4 \pi} \; \nu (k) \de k \; .
\ee

In terms of this distribution function, the conserved quantities can be written as
\bea
    M & = & N \int \frac{\de k}{4 \pi} \; \nu (k),
    \label{HSM-M} \\
    P & = & N \int \frac{\de k}{4 \pi} \; \nu (k) \; k,
    \label{HSM-P} \\
    E & = & E_0 + N \int \frac{\de k}{4 \pi} \; \nu (k) \; \frac{k^2}{2},
    \label{HSM-E}
\eea
where the momentum is defined only modulo $2\pi$.
From now on, we will drop the constant energy shift $E_0$.

In a hydrodynamic description we assume a distribution of the uniform type
\be
   \nu(k) = \left\{ \begin{array}{cl}
   1, & \qquad{\rm if \;} -\pi< k_L < k < k_R<\pi,
  \\
   0, & \qquad{\rm otherwise},
   \end{array} \right.
 \label{tildenuCO}
\ee
where $k_{R,L}$ are some numbers.
Using (\ref{tildenuCO}) and introducing space dependent fields instead of constants we write (\ref{HSM-M},\ref{HSM-P}) as
\bea
   M & = &
   \int \de x\, \frac{k_R - k_L}{4 \pi} =\int \de x \,\rho, \\
   P & = &
   \int \de x\, \frac{k_R^2 - k_L^2}{8 \pi} = \int \de x\, \rho \, v,
\eea
which suggests the identification
\be
   k_{R,L} = v \pm 2 \pi \rho \; .
 \la{HSMident}
\ee
Then the hydrodynamic Hamiltonian follows from (\ref{HSM-E}):
\bea
   H_{\rm HSM} & = &
    \int \de x\, \frac{k_R^3 - k_L^3}{24 \pi}
    =
    \int \de x \left[\frac{1}{2} \; \rho \; v^2
   + \frac{2}{3} \; \pi^2 \; \rho^3 \right] \; ,
   \label{H-HSM}
\eea which corresponds, as expected, to the (gradientless) hydrodynamic of a $\lambda =1$ spin-less Calogero-Sutherland model (\ref{spingless}).

We think of slowly varying fields $\rho(x,t)$ and $v(x,t)$ as of classical fields obeying the Poisson relation $\left\{\rho(x), v(y)\right\}=\delta^{\prime}(x-y)$. Then (\ref{H-HSM}) generates the evolutions equations
\begin{eqnarray}
	\dot{\rho} & = & -\partial_{x}\left(\rho v \right),
 \nonumber \\
	\dot{v} & = & -\partial_{x}\left( \frac{v^{2}}{2}
	+\frac{\pi^{2}}{2}4 \rho^{2}\right).
 \la{HShydro2}
\end{eqnarray}
One can easily recognize in (\ref{HShydro2}) the hydrodynamics of spinless Calogero-Sutherland model (\ref{spingless}) for $\lambda=1$. The correspondence between eigenstates and eigenenergies of Haldane-Shastry model with $\lambda=2$ spinless Calogero-Sutherland model has been noticed in the original paper \cite{1988-Haldane-HS}. The degeneracy of the states due to the $SU(2)$ invariance and Yangian symmetry is lost in our classical hydrodynamics model.

For comparisons with the derivations from freezing trick \cite{1993-Polychronakos} in section \ref{sec:Lattice} we express (\ref{H-HSM},\ref{HShydro}) in terms of $\rho_{s}$ and $v_{s}$ used in the main body of the paper. We identify the density $\rho=M/N=\rho_{\downarrow}$ as the density of spin-down particles and the velocity $v$ as a velocity of spin-down particles relative to the static background of spin-up  particles, i.e., $v=v_{\uparrow}-v_{\downarrow}=-2v_{s}$. The charge density corresponding to the lattice with spacing one is just $\rho_{0}=1$. We summarize
\be
   \rho = \rho_{\downarrow}=\frac{\rho_0 -\rho_{s}}{2} \, , \qquad \qquad
   v = - 2v_{s}\, , \qquad \qquad  \rho_{0}=1.
 \la{frid}
\ee
Using (\ref{frid}) we rewrite (\ref{H-HSM}) as
\begin{equation}
	H_{\rm HSM}=\int \de x\left\{\rho_{0}v_{s}^{2}-\rho_{s}v_{s}^{2}
	 +\frac{\pi^{2}\rho_{0}\rho_{s}^{2}}{4}-\frac{\pi^{2}\rho_{s}^{3}}{12}\right\} \; ,
 \label{eq:haldane-shastry hydrodynamics}
\end{equation}
where we neglected a constant and a term linear in $\rho_s$, which amounts to a shift in the chemical potential. The evolution equations for the spin density and spin velocity follow from (\ref{HShydro2},\ref{frid})
\begin{eqnarray}
	\dot{\rho}_{s} & = & -\partial_{x}\left\{ 2v_{s}\rho_{0}-2v_{s}\rho_{s}\right\} \; ,
 \nonumber \\
	\dot{v}_{s} & = & -\partial_{x}\left\{ -v_{s}^{2}
	+\frac{\pi^{2}}{2}\rho_{0}\rho_{s}-\frac{\pi^{2}}{4}\rho_{s}^{2}\right\} \; .
 \label{HShydro}
\end{eqnarray}

We notice that the above (\ref{eq:haldane-shastry hydrodynamics}, \ref{HShydro}) is nothing but the strong interaction limit of the sCM (\ref{eq:CO-case-freezing},\ref{eq:o1rhos},\ref{eq:o1vs}).

Finally, we remark that it is easy to check that the distribution function (\ref{tildenuCO}) implies that $-\frac{\pi\rho_{s}}{2}\leq v_{s}\leq\frac{\pi\rho_{s}}{2}$ and therefore corresponds to the CO regime of spin-Calogero model.

Both in this appendix and in writing classical hydrodynamics for sCM we neglected the degeneracy of the corresponding quantum models due to the Yangian symmetry \cite{1992-HaldaneEtAl,HaHaldane93}. We assumed that during the evolution string states are not excited. Of course, the degeneracy plays a very important role for perturbed integrable systems and for the hydrodynamics at finite temperatures.




\begin{thebibliography}{15}
\bibitem{linear-note}
 	This is a peculiar property of the  spin-Calogero model and, moreover of CO regime. In PO regime this property does not hold, see appendix \ref{app:allcases} for details.
\bibitem{monotonicity}
	The function $k(\kappa)$ is monotonic and,  therefore, $\sgn(k-k')=\sgn(\kappa-\kappa')$.
	This trick is specific for Calogero-type models with scattering phase given by $\sgn(k-k')$ and works for Haldane-Shastry model in the absence of umklapps.
\end{thebibliography}
\end{document}